\documentclass[twocolumn,showpacs,
  aps,superscriptaddress,
  prd,notitlepage,showkeys,
  nofootinbib,10pt]{revtex4-2}
\usepackage[normalem]{ulem}
\usepackage{amssymb}
\usepackage{amsmath}
\usepackage{graphicx}
\usepackage{dcolumn}
\usepackage[colorlinks,urlcolor=blue,citecolor=blue,linkcolor=blue]{hyperref}
\usepackage[dvipsnames]{xcolor} 
\usepackage{lineno}
\usepackage{subfigure}
\usepackage{braket}
\usepackage{physics}
\usepackage{orcidlink}
\usepackage{cuted}

\newcommand{\GSSI}{Gran Sasso Science Institute (GSSI), I-67100 L’Aquila, Italy}
\newcommand{\GranSasso}{INFN, Laboratori Nazionali del Gran Sasso, I-67100 Assergi, Italy}

\begin{document}

\newcommand{\IUCAA}{Inter-University Centre for Astronomy and
  Astrophysics, Post Bag 4, Ganeshkhind, Pune - 411007, India}

\newcommand{\ICTS}{
International Centre for Theoretical Sciences, Tata Institute of Fundamental Research, Bangalore 560089, India}
\newcommand{\WSU}{Department of Physics and Astronomy, Washington State University, 1245 Webster, Pullman, Washington 99164-2814, U.S.A.}

\title{Binary black holes in the heat of merger}

\author{Samanwaya Mukherjee\orcidlink{0000-0002-9055-5784}}
\email{samanwaya.physics@gmail.com}
\affiliation{\IUCAA}
\affiliation{\ICTS}

\author{Sayak Datta\orcidlink{0000-0002-4774-0298}}
\email{sayak.datta@gssi.it}
\affiliation{\GSSI}
\affiliation{\GranSasso}

\author{Sukanta Bose}
\email{sukanta@wsu.edu}
\affiliation{\WSU}

\author{Khun Sang Phukon \orcidlink{0000-0003-1561-0760}}
\email{k.s.phukon@bham.ac.uk}
\affiliation{School of Physics and Astronomy and Institute for Gravitational Wave Astronomy,\\University of Birmingham, Edgbaston, Birmingham, B15 2TT, United Kingdom}


\begin{abstract}
   A black hole binary approaching merger undergoes changes in its inspiral rate as energy and angular momentum are lost from the orbits into the horizons. This effect strengthens as the black holes come closer. We use numerical relativity data to model this so-called tidal heating in the strong gravity regime. We also present a frequency-domain approximant for nonspinning black hole binaries that accounts for tidal heating effects up to the merger frequency. The approximant includes horizon parameters that characterize the nature of the compact objects. This model serves two main purposes: (1) leveraging the stronger effects of tidal heating near merger, it allows for more robust tests for the presence of black holes compared to the tests with inspiral-only analytical waveforms, and (2) by applying this model to a binary black hole baseline waveform that incorporates tidal heating, one can construct more accurate point-particle waveforms free from the finite-size effects of the component objects. We discuss its ramifications in modeling binary neutron star systems.
\end{abstract}

\maketitle

\section{Introduction} In an inspiralling binary of compact objects, parts of the energy and angular momentum of the orbit are drained due to viscous dissipation (material objects) or absorption by the horizons (black holes). This effect is strong for black holes (BHs) as their horizons define one-way hypersurfaces that do not allow any mass-energy to escape outward. A BH in a binary experiences this due to the dynamical tidal effects of its companion on its horizon geometry~\cite{Hartle:1973zz, Vega2011,OSullivan:2014ywd,OSullivan:2015lni, Prasad:2021dfr}. The exchange of energy and angular momentum back-reacts on the orbit and alters the description of the binary's gravitational-wave (GW) phase~\cite{datta2020recognizing,Mukherjee:2022wws}. This phenomenon causes the horizons to grow in area and entropy~\cite{HawkingHartle} -- a dissipative process aptly termed \emph{tidal heating} (TH)~\cite{Hartle:1973zz, Hughes:2001jr}.
\footnote{In this work, however, we consider scenarios in which such dissipation may also be present, at least partially, in non-black-hole compact objects. Accordingly, we use the term ``tidal heating" to broadly refer to this phenomenon in generic compact binaries that may or may not be BBHs.}
This effect is especially pronounced in extreme-mass-ratio inspirals (EMRIs), where it introduces thousands of orbital cycles~\cite{Hughes:2001jr, Taracchini:2013wfa, Harms:2014dqa, Datta:2019euh, Datta:2019epe, Datta:2024vll}, which space-based detectors like LISA are designed to observe~\cite{Colpi:2024xhw}. For stellar to intermediate-mass binaries observable by the LIGO-Virgo and KAGRA network~\cite{TheLIGOScientific:2014jea,VIRGO:2014yos,Aasi:2013wya} and the proposed third-generation ground-based GW detectors~\cite{Reitze:2019iox,Abac:2025saz}, the total accumulated dephasing due to horizon fluxes is smaller, but can become important for binaries with high component spins, asymmetric masses, or large signal-to-noise ratios (SNRs)~\cite{Mukherjee:2022wws}. Our work focuses on such comparable-mass binaries with total masses ranging from a few to a few hundred solar masses.

Tidal heating affects binaries both with and without black holes, although its impact depends on the nature of the inspiraling objects. 
For the theoretical exotic compact objects (ECOs) that can act as BH-mimickers~\cite{Cardoso:2019rvt}, TH effects depend on the location of their surfaces~\cite{Maselli:2017cmm, Datta:2019epe, Datta:2020rvo,Maggio:2021uge}, and are different from that of BHs due to a finite reflectivity.  For a binary of generic compact objects, the amount of TH flux lost from the orbit can be expressed as a fraction of the flux that would be lost if it were a BBH system of the same component masses and spins. One can use the \textit{horizon parameter} ($H$)~\cite{Maselli:2017cmm, Datta:2019epe,Mukherjee:2022wws} to represent this fraction. $H$ takes values between 0 and 1, the latter representing a BH. An object with $H\sim 0$ partakes in negligible TH. One such example is a neutron star (NS) whose viscous dissipation of energy from the orbits, under the standard assumption of its interior composition, is about 10 orders of magnitude smaller than a black hole of the same mass~\cite{datta2020recognizing,Glampedakis:2013jya}. One can safely assume, therefore, that NSs have $H=0$.

The conventional assumption that the compact objects observed by LIGO-Virgo-KAGRA are either NSs or BHs necessitates adopting a choice of distinction between these objects based on their intrinsic properties. The maximum mass that an NS can support against gravity depends on the equation of state (EoS) describing its internal matter and, 
taking into account the existing EoS models, has an approximate upper bound~\cite{Rhoades:1974fn, Kalogera:1996ci}. In GW data analysis, this limit is loosely set at $3M_\odot$~\cite{KAGRA:2021vkt}, above which a compact object is typically classified as a BH. 
However, if one entertains the possible existence of exotic objects heavier than NSs mimicking BHs in a binary coalescence~\cite{Cardoso:2019rvt}, mass estimates alone may be insufficient for their identification. Additionally, classification of compact objects in the mass-gap range ($\sim$2.5--5$M_\odot$)~\cite{Bailyn:1997xt,Ozel:2010su,Farr:2010tu,Kreidberg:2012ud} becomes even more uncertain~\cite{LIGOScientific:2020zkf,LIGOScientific:2021usb,LIGOScientific:2021usb,KAGRA:2021vkt,KAGRA:2021duu,LIGOScientific:2024elc},
as theoretical predictions and observational constraints are insufficient to categorically distinguish between BHs and NSs. In such cases, tests involving $H$ offer a promising avenue~\cite{datta2020recognizing}.

The strength of TH changes as a compact binary evolves. At later stages of the inspiral-merger phase, its effects are more significant due to the stronger tidal fields. This was discussed by Scheel \textit{et al.}~\cite{Scheel:2014ina} using the evolution of the Christodoulou mass of BHs in numerical relativity (NR) data from the SXS catalog~\cite{Boyle:2019kee,Scheel:2025jct}. Consequently, a complete gravitational waveform containing \emph{tunable} TH effects up to the merger would be a more powerful BH identifier compared to inspiral-only waveforms usually truncated at the innermost stable circular orbit (ISCO).
Moreover, a complete waveform approximant for TH also helps refine the \emph{point-particle} model, one that is devoid of any finite-size effect of the binary components, and differs from a BBH waveform in that the latter has TH effects. 
A better point-particle model may serve as a better baseline for constructing, e.g., binary neutron star (BNS) waveform models, by introducing in them the tidal deformability effects~\cite{Dietrich:2017aum,Dietrich:2019kaq,Abac:2023ujg}.
Currently the baselines used for this purpose are BBH waveforms that inherently contain, through NR calibrations, contributions from horizon fluxes.
The improvement in waveform systematics achieved by correcting for the horizon fluxes could be significant, especially for the next-generation detectors where the SNRs will be large and waveform systematics will dominate over the statistical errors~\cite{Purrer:2019jcp}.
In this work, we broadly address the following questions: Can the enhanced horizon fluxes in NR data make a significant difference in GW data analysis? Can they reinforce tests of the black hole nature of the components of a binary? If ignored, can they contaminate BNS waveform models? Since post-Newtonian (PN) expansions lose accuracy in the strong-gravity regime, we address these questions using publicly available NR data from SXS, and find affirmative answers. Effects of TH have been largely neglected in the analyses of binaries detected by LIGO and Virgo due to their weak contribution in the inspiral phase. However, ignoring this effect in the strong-gravity regime may have notable consequences.

As a first step in this effort, we construct, for the first time, a model of TH for nonspinning BBH systems valid up to the merger. This model lays the groundwork for building accurate point-particle baselines and for developing complete waveform models of BNS systems, or binary compact objects with arbitrary TH.

The paper is organized as follows: in Section~\ref{sec: horizon flux in NR} we briefly describe the NR data used in this work. In Section~\ref{sec:mdot model} we build a model of the horizon fluxes that combines PN and NR information and remains valid up to the merger frequency. In Section~\ref{sec:phase approx} we use this model of the fluxes to create a frequency-domain model of the GW phase for constructing gravitational waveforms. Such waveforms represent generic binaries with arbitrary TH fluxes when tunable horizon parameters are introduced; these are constructed in Section~\ref{sec:Waveforms for non-black-hole compact objects}.
In Section~\ref{sec:bns} we explore the implications of the model presented here in modeling binary neutron star waveforms. Section~\ref{sec:conclusion} summarizes the work and discusses its ramifications in gravitational waveform modeling and future directions.

\section{Horizon fluxes from NR simulations}\label{sec: horizon flux in NR}

For each simulation of BBHs, SXS provides data for the properties of the apparent horizons, tracking their evolution up to the formation of a common horizon~\cite{Prasad:2020xgr}. The data contains evolution of the Christodoulou mass, horizon area and the spin of each BH. Time (in units of the total mass $M$) when the common apparent horizon starts to form is also specified, and denotes the moment of merger of the BHs. We utilize these data for modeling horizon flux of BHs from the late inspiral to the merger regime. Detailed analyses of the geometric aspects and fluxes associated with the dynamical horizons during binary black hole mergers have been carried out in Refs~\cite{Pook-Kolb:2020zhm,Pook-Kolb:2020jlr}.
These studies focus on the geometric properties of individual simulations, primarily for head-on collisions. In contrast, our work presents a systematic study of the horizon-absorbed fluxes for quasicircular nonspinning simulations, and compares them directly with the GW flux radiated to infinity.

\begin{figure}[ht]
    \centering
    \includegraphics[width=\linewidth]{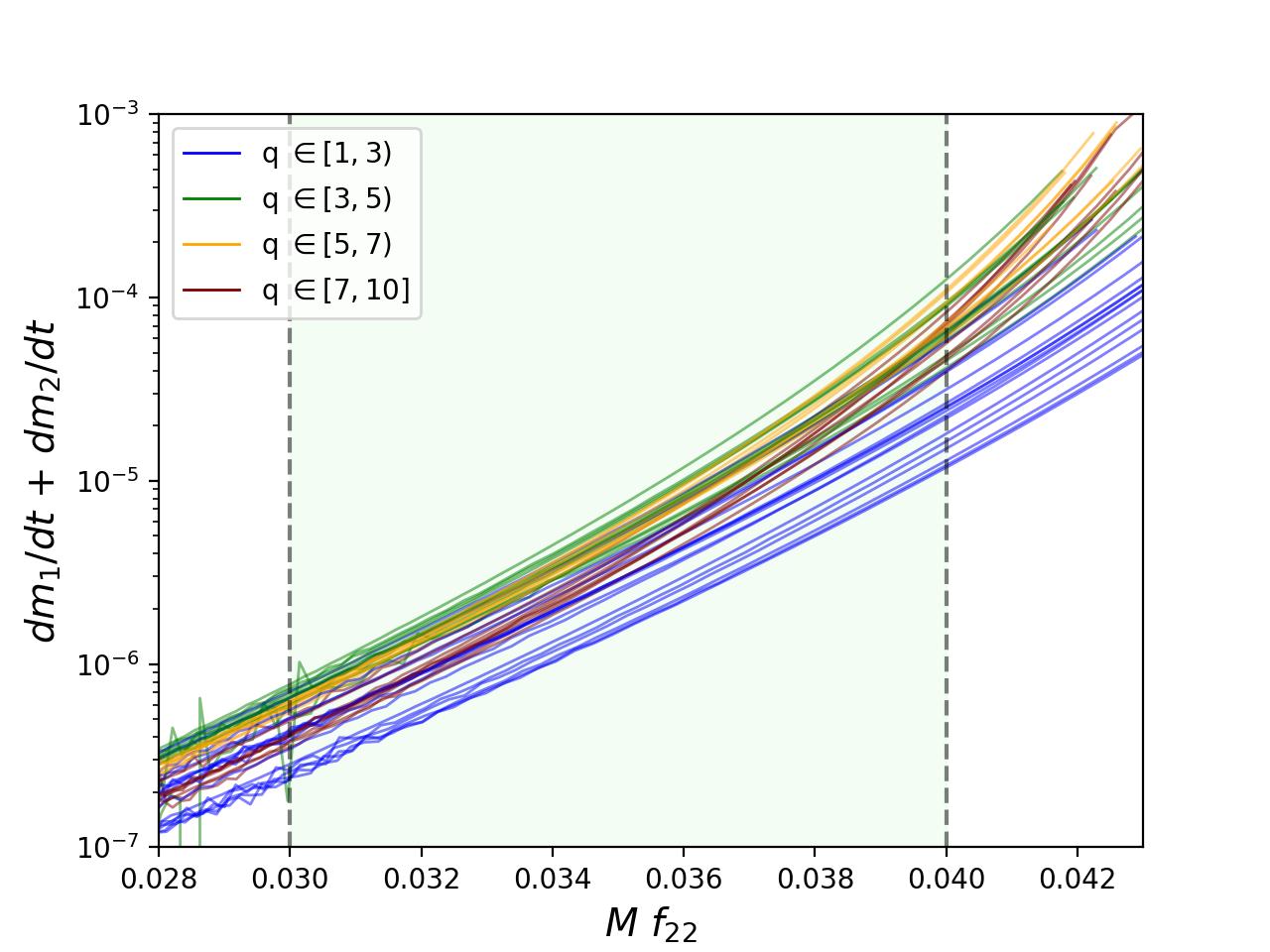}
    \caption{$\dd M/\dd t$ as a function of dimensionless frequency of the $(2,2)$ mode, from 50 nonspinning SXS simulations. The green shaded region is used for modeling.
    }
    \label{fig:dmdt_all}
\end{figure}

\begin{figure*}[ht]
    \centering
    \includegraphics[width=\linewidth]{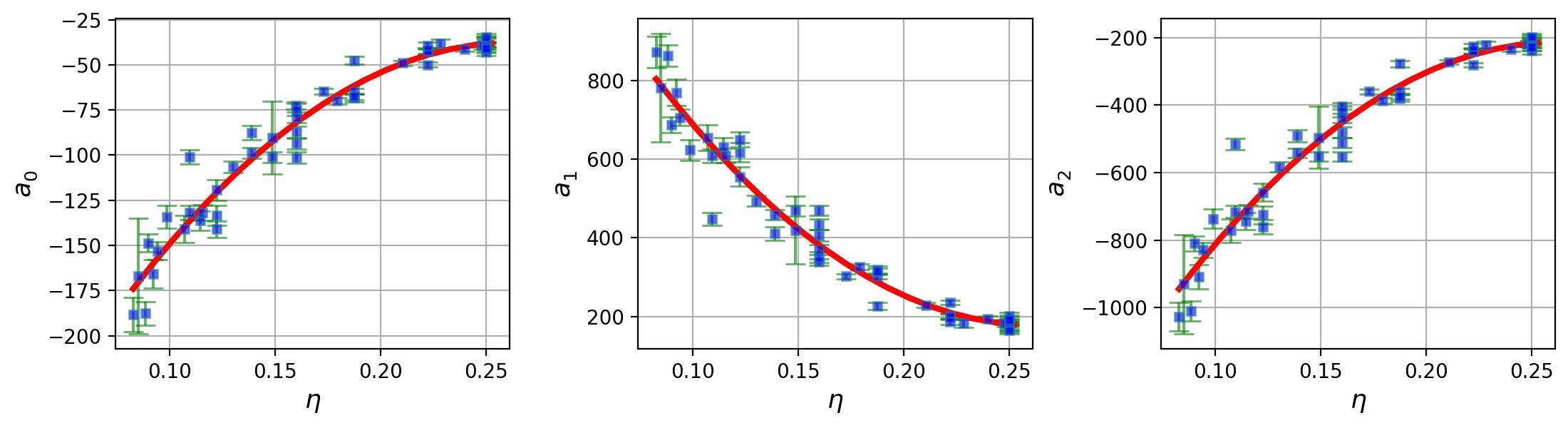}
    \caption{Data (blue dots) and best fits (red curves) for $a_i$ introduced in Eq.~\eqref{eq:mdotNR}, using 50 BBH SXS simulations. 1-$\sigma$ error bars are shown.
    }
    \label{fig:a params}
\end{figure*}

We perform modeling of nonspinning BH binaries with individual masses $m_1$ and $m_2$, total mass $M=m_1+m_2$, and the symmetric mass ratio $\eta=m_1m_2/M^2$. 
Magnitude of the rate of change of BH masses ($\dot{m_i}=\dd m_i/\dd t$, $i=1,2$) is extremely small at lower frequencies, and the horizon data from the highest resolution available start to resolve $\dot{m_i}$ only after $Mf\sim 0.03$, $f$ being the GW frequency of the $(2,2)$ mode, labeled as $f_{22}$ in the figures. Figure~\ref{fig:dmdt_all} shows the mass evolution from 50 nonspinning SXS simulations used in this study, demonstrating adequate resolution in the frequency range of interest.

\section{Inspiral-merger model of horizon flux}\label{sec:mdot model} 

\begin{figure*}[ht]
\centering
 \subfigure[]{{\label{fig:flux_q1}}\includegraphics[width=0.3297\textwidth]{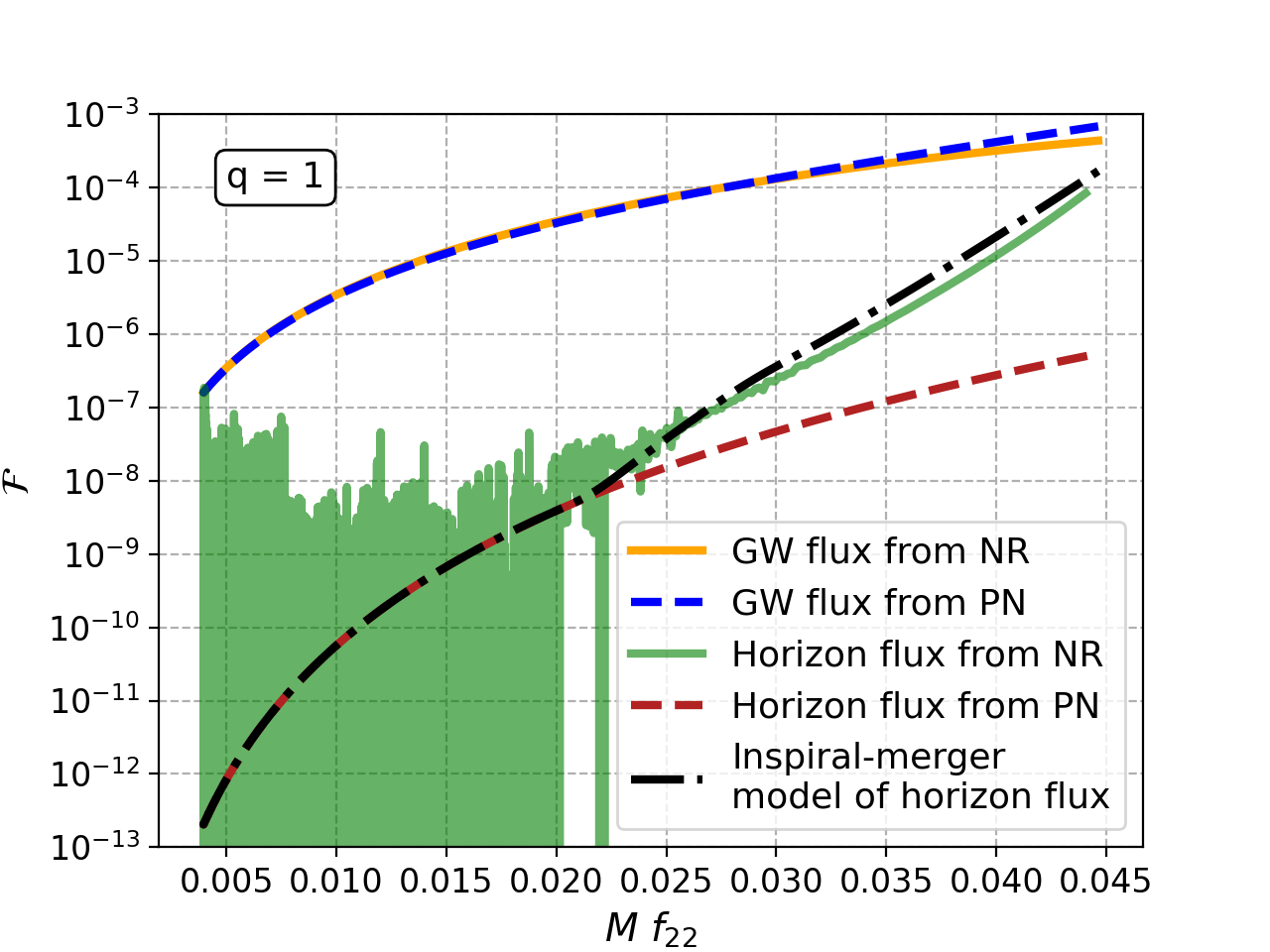}}
 \hfill
 \subfigure[]{{\label{fig:flux_q4}}\includegraphics[width=0.3297\textwidth]{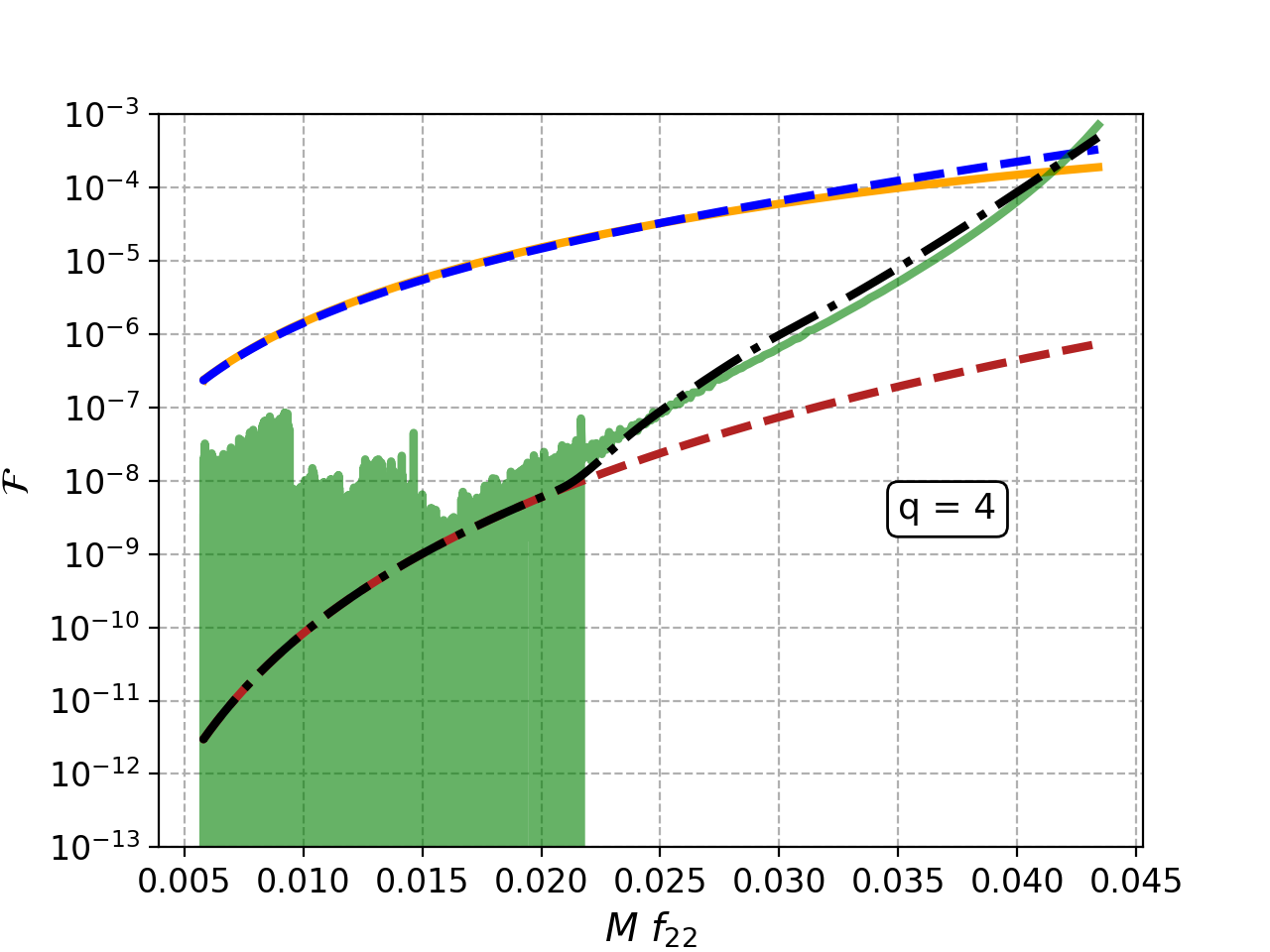}}
 \hfill
 \subfigure[]{{\label{fig:flux_q10}}\includegraphics[width=0.3297\textwidth]{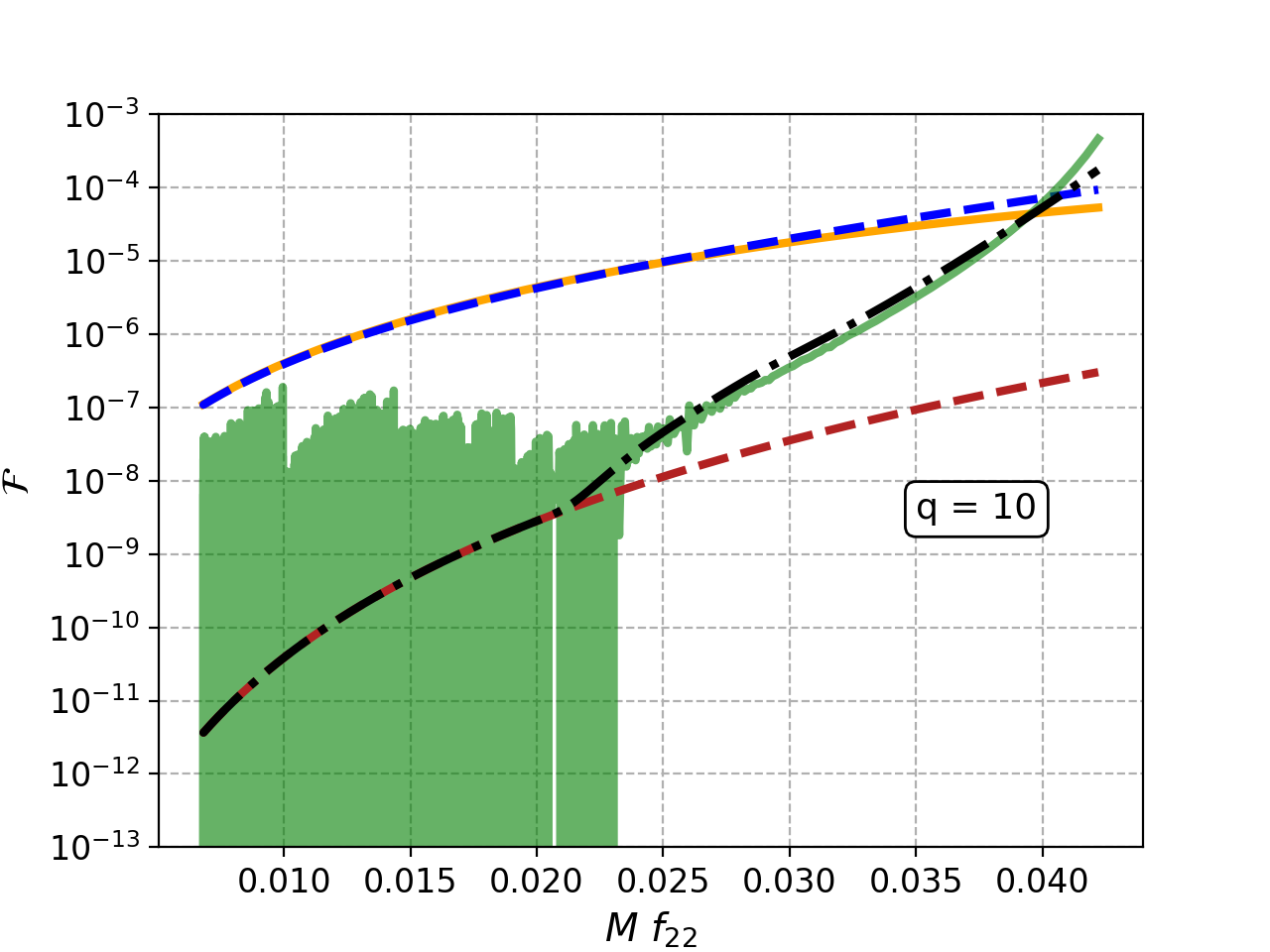}}
 \caption{Comparison of the energy flux radiated to future null infinity and absorbed by the black hole horizons, as computed from the SXS catalog of numerical relativity simulations and from current state-of-the-art post-Newtonian expressions. The black dashed-dotted lines show the inspiral-merger model of the horizon-absorbed fluxes by fitting them with three phenomenological parameters within $0.03\leqslant Mf_{22}\leqslant 0.04$ and connecting them to the PN expression using a Planck taper. }
\label{fig:flux-comparison}
 \end{figure*}

We model the rate of the total mass-energy absorption by the two BHs ($\dot{M}=\dot{m_1}+\dot{m_2}$) using the ansatz
\begin{equation}\label{eq:mdotNR}
    \log(\dot{M}^{}_{\rm NR})=\log(\dot{M}^{}_{\rm PN})\left\{1+\eta\sum_{i=0}^2a_iv^{i+1}\right\}\,,
\end{equation}
where $\dot{M}^{}_{\rm PN}$ is taken from the most recent PN expressions derived in Ref.~\cite{Saketh:2022xjb}. $v=(\pi Mf)^{1/3}$ is the PN expansion parameter.
We note that a traditional PN expansion of dynamical quantities in powers of $v$ is expected to break down close to merger, rendering analytical calculations difficult. In this regime, the binary evolution, including tidal dynamics, becomes highly nonlinear.
The assumed form of the horizon fluxes in Eq.~\eqref{eq:mdotNR} is purely phenomenological in nature, designed to model the NR data as accurately as possible.
Using the logarithm of the fluxes allows for a more accurate representation of their steep increase near the merger, while the chosen ansatz ensures that the PN behavior is correctly recovered at low frequencies.
 We fit the NR data of horizon fluxes within the frequency range $0.03\leqslant Mf\leqslant 0.04$ for each of the 50 simulations. The upper frequency for fitting is chosen so that we have resolved data for all the simulations, since binaries with different mass ratios form a common horizon at different frequencies.
The phenomenological parameters $\{a_0,a_1,a_2\}$ are evaluated for each simulation, and are fitted with a second-order polynomial in $\eta$: $a_i=\sum_{j=0}^{2}\alpha_{ij}\eta^{j}$. The matrix $\alpha$ is given by
\begin{equation}
    \alpha = \begin{pmatrix}
                -328.06 &  2210.473 & -4205.79 \\
                1510.162 & -10140.382 & 19288.622\\
                -1768.95 & 11818.964 & -22445.763
                \end{pmatrix}\,,
\end{equation}
and the $j$ in $\eta^j$ is an exponent.

Figure~\ref{fig:a params} shows the numerical values and the best fits for these parameters. To construct a complete inspiral-merger (IM) model, we smoothly transition between the PN and NR fluxes using a Planck taper function $w$, defining 
\begin{equation}
    \dot{M}_{\rm IM} = (1 - w)\dot{M}_{\rm PN} + w\dot{M}_{\rm NR}\,,
\end{equation}
with $w$ increasing from 0 to 1 in the interval $0.019 \leqslant Mf \leqslant 0.032$. Beyond $Mf=0.04$, $\dot{M}_{\rm NR}$ is extrapolated up to the merger frequency. 1-$\sigma$ error bars are shown corresponding to every NR simulation used in the modeling, in estimating each of the phenomenological parameters.

In Fig.~\ref{fig:flux-comparison}, we present the total energy flux into the horizons of BHs and the flux radiated to future null infinity, as obtained from three different NR simulations. In the same figure, we also compare these numerical fluxes with their analytical counterparts computed from PN calculations~\cite{Blanchet:2023bwj,Saketh:2022xjb}, and show the IM model of horizon fluxes described above. We calculate the GW flux that escapes to infinity from the formula~\cite{Boyle:2008ge}
\begin{equation}\label{eq:edot_infty}
    \dot{E}_{\rm \infty}=\frac{1}{16\pi}\sum_{l=2}^8\sum_{m=-l}^l|r\dot{h}_{lm}|^2\,,
\end{equation}
$h_{lm}$ being the $(l,m)$ mode of GW strain that is extracted at a radius $r$. The sum over $l$ is taken up to 8 as the NR simulations contain modes up to that value. 
We find that while the flux to infinity predicted by PN theory reasonably agrees with NR, the horizon fluxes diverge significantly once NR begins to resolve them. In all cases, PN theory substantially underestimates TH, with the discrepancy growing at higher mass ratios. 
With $M=90 M_\odot$ and $1\leqslant q \leqslant 10$, the NR-informed model of the TH fluxes improves the mismatches between waveforms with and without TH by at least an order of magnitude (see Sec.~\ref{sec:Waveforms for non-black-hole compact objects} for a discussion on mismatches) as compared to the analytical PN expression extrapolated up to the merger frequency. 

Remarkably, for binaries with $q \gtrsim 3$, the NR-reported horizon flux exceeds the radiated flux to infinity close to the merger, as shown for $q=4$ and $q=10$ in Figs.~\ref{fig:flux_q4} and~\ref{fig:flux_q10}. In the literature, studies of horizon dynamics and comparisons between the infalling energy flux at BH horizons and the outgoing GW flux have been performed in the context of dynamical horizons~\cite{Gupta:2018znn}. However, to the best of our knowledge, the striking fundamental feature of BBH coalescences that horizon flux exceeds GW flux near merger for $q\gtrsim 3$ has not been previously reported.

\section{Inspiral-merger phase approximant for tidal heating}\label{sec:phase approx}

We assume that both the component objects are of the same kind -- black hole or otherwise -- so the total amount of TH in a binary can be parametrized by a single horizon parameter $H$. This parameter is the ratio of the absorbed TH fluxes in a binary of generic compact objects to those in a BBH, and is introduced as a free parameter independent of the objects' intrinsic properties.
In our treatment, $H$ is a frequency-independent intrinsic parameter that is related to the reflectivity of the compact objects.
To find the phase correction due to TH in the frequency domain, we use the energy balance condition $-\dot{E}_{\rm orb} = \dot{E}_{\rm \infty} + H\dot{M}$,
where $\dot{E}_{\rm orb}$ and $\dot{E}_{\rm \infty}$ are the time rates of change of the orbital energy and the GW flux to infinity, respectively, and horizon parameter $H$ is as defined earlier. For orbital binding energy $E_{\rm orb}(v)$, we use 4PN-accurate expression~\cite{Blanchet:2023bwj}.

Under the stationary phase approximation (SPA) \cite{Tichy:1999pv}, the frequency-domain phase for a compact binary system with horizon parameter $H$ can be written as 
\begin{equation}
\begin{aligned}\label{eq:psi_v}
    \psi(v,H)=~&2(t_c/M)v^3-2\phi_c-\frac{\pi}{4} \\& -\frac{2}{M}\int_{v_i}^v \frac{E_{\rm orb}'(\bar{v})}{F(\bar{v},H)}(v^3-\bar{v}^3)\dd \bar{v}\,,
\end{aligned}
\end{equation}
where $F(v)=-\dot{E}_{\rm orb}(v)$. 
Here $t_c$ and $\phi_c$ capture the arbitrary time and phase shift freedom on the complete waveform, respectively. We  use the IM model of $\dot{M}$ constructed earlier, and compute the frequency-domain phase using Eq.~\eqref{eq:psi_v} numerically within the frequency range $0.0035\leqslant Mf\leqslant 0.04$. The phase difference between a BBH system and a binary of ``point particles" can be written as 
\begin{equation}\label{eq:delta_psi_im}
    \delta\psi_{\rm TH,\, BBH}^{\rm (IM)}=\psi(v,H=1)-\psi(v,H=0)\,.
\end{equation} 

\begin{figure}[h]
    \centering
    \includegraphics[width=\linewidth]{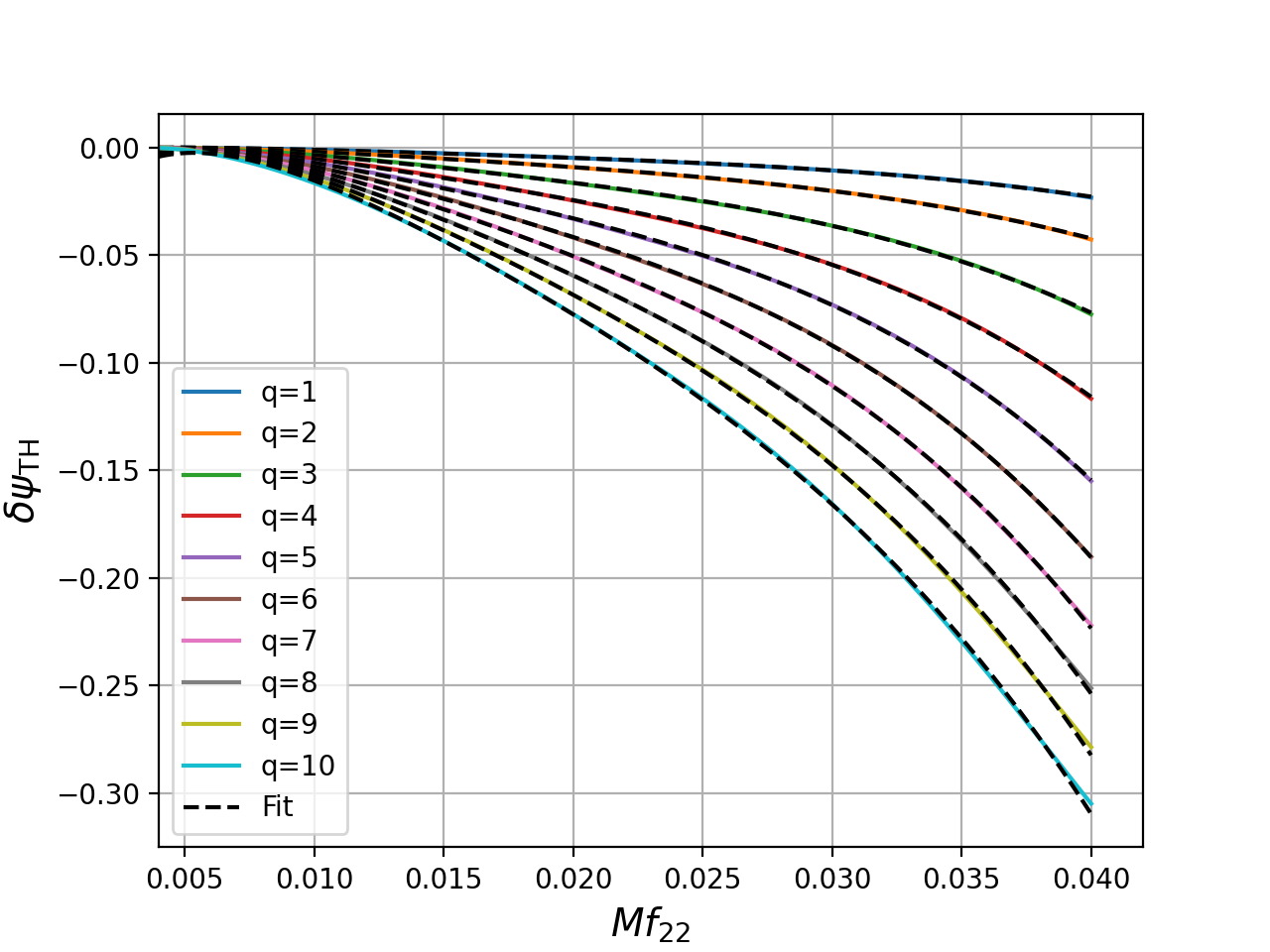}
    \caption{Phase contribution of TH of BBHs calculated numerically (solid curves) from Eqs.~\eqref{eq:psi_v} and~\eqref{eq:delta_psi_im}, and the fitted model using Eq.~\eqref{eq:fit ansatz} (black dashed curves).}
    \label{fig:psi model}
\end{figure}

$\delta\psi_{\rm TH,\,BBH}^{\rm (IM)}$ can be generated for any values of $m_1$ and $m_2$ by numerically computing it using Eqs.~\eqref{eq:psi_v} and~\eqref{eq:delta_psi_im} dynamically while generating a waveform. To aid a faster waveform generation, we construct a fitted model that can be used directly instead of performing the integration dynamically. We compute numerical $\delta\psi_{\rm TH}^{\rm (IM)}$ for 10 mass-ratio values ranging from 1 to 10 within the frequency range $0.0035\leqslant Mf \leqslant 0.04$, and model the numerical phase with 4 phenomenological coefficients $\{b_i\},i=0,1,2,3$:
\begin{equation}\label{eq:fit ansatz}
    \delta\psi_{\rm TH,\,BBH}^{\rm (IM)} = \delta\psi_{\rm TH,\,BBH}^{\rm (PN)}+\frac{3}{128\eta v^5} \sum_{i=0}^3b_iv^{i+9}, 
\end{equation}
where the coefficients $b_i$ are fitted with a third-order polynomial in $\eta$ as $b_i=\sum_{j=0}^{3}\beta_{ij}\eta^{j}$.
The matrix $\beta$ we find to be
\begin{equation}
\resizebox{.47\textwidth}{!}{$
\beta=\left(
\begin{array}{cccc}
        -85.689 & 6787.851 & -43705.198 & 75611.031\\
        723.846 & -57045.812 & 367908.088 & -637826.407\\
        -2898.275 & 160880.802 & -1017873.869 & 1764975.530\\
        3318.419 & -148686.91 & 925937.443 & -1605648.237\end{array}
\right)\,.
$}
\end{equation}
Figure~\ref{fig:psi model} shows the numerical phase and the fitted model, up to a mass ratio $q=10$.

\section{Waveforms for non-black-hole compact objects}\label{sec:Waveforms for non-black-hole compact objects} 

TH affects the binary evolution only up to the frequency at which the objects merge, which entails defining a cutoff condition based on the frequency at merger.
We use the phenomenological fit of the peak frequency $f^{\rm peak}$ from Ref.~\cite{Healy:2017mvh}. In the comparable mass-ratio case, the fit is given by,
\begin{equation}
    M\omega_{(2,2)}^{\rm peak}=W_0+W_2 (\delta M)^2+W_4 (\delta M)^4\,,
\end{equation}
with $\omega_{(2,2)}=2\pi f^{\rm GW}_{(2,2)}$ (in the present context, simply $2\pi f$), and $\delta M=(m_1-m_2)/(m_1+m_2)$.  The fit parameters are $W_0=0.3586$, $W_2=-0.121$, and $W_4=0.0431$.
We taper the approximant starting from $1.15f^{\rm peak}$ and ending at $1.35f^{\rm peak}$ using a Planck taper window $w$~\cite{Mukherjee:2023pge}.

To create binary waveforms with arbitrary TH, the frequency-domain approximant presented in Eq.~\eqref{eq:fit ansatz} can be used on a waveform of BBH system that includes TH up to the merger frequency. We use \texttt{IMRPhenomD\_Horizon}~\cite{Mukherjee:2023pge} as this BBH baseline. We note that for a binary of compact objects (BCO) similar in nature, with horizon parameter $H$, the phase correction due to TH becomes
\begin{equation}
    \delta\psi^{\rm(BCO)}_{\rm TH}=H\delta\psi^{\rm(BBH)}_{\rm TH}\,,
\end{equation}
where we have added a superscript for BBHs to remind the reader that the $\delta\psi_{\rm TH}$ in Eq.~\eqref{eq:delta_psi_im} represents TH in BBH systems, to avoid ambiguity. So, the frequency-domain waveform becomes
\begin{equation}\label{eq:hf bco}
    \tilde{h}^{\rm(BCO)}(f)=\tilde{h}^{\rm(BBH)}(f)\exp\left\{i(1-H)\delta\psi^{\rm(BBH)}_{\rm TH}\right\}\,,
\end{equation}
with the convention that $ \tilde{h}(f)=\tilde{A}(f)e^{-i\psi(f)}$. We note that the high-frequency cutoff for the phase correction is defined considering BBHs only, and non-BH compact objects would merge at a lower frequency due to their smaller compactness. Consequently, termination of the phase correction should ideally depend on the compact object being considered \cite{Ghosh:2025wex}. This is a limitation of this model that can be rectified only in a model-dependent manner.

\begin{figure}
    \centering
    \includegraphics[width=\linewidth]{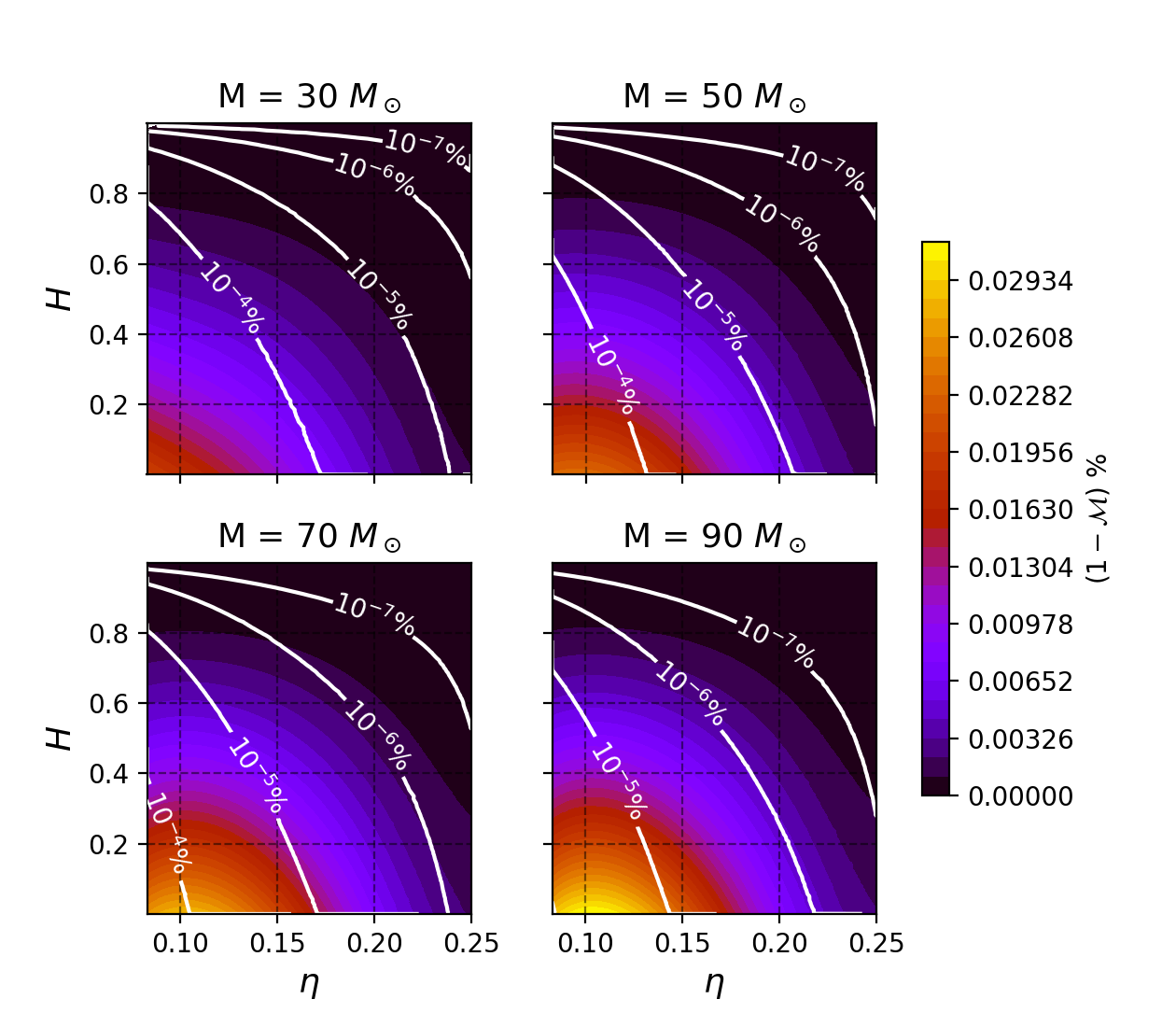}
    \caption{Percentage mismatches ($1-\mathcal{M}$) between the full BBH waveforms and BCO waveforms  (Eq.~\eqref{eq:hf bco}) in A+ design sensitivity curve of LIGO
    plotted against $H$ and $\eta$, and represented by the colorbar. Four different total mass values are considered, and the initial frequency for match calculation is fixed at 20~Hz. For the colored contours, the upper cutoff frequency for the mismatch is taken to be the BBH merger frequency or the Nyquist frequency, whichever is smaller. The overlaid white contours indicate mismatches between the same waveforms, but computed only up to the ISCO frequency -- with the corresponding mismatch values labeled on the contours.}
    \label{fig:mismatch_H}
\end{figure}

In this work we build a model for binaries of compact objects without accounting for their compactness or internal composition. Our model differs from a standard BBH model only by including a parameterized description of TH experienced by the objects prior to merger. Extending this framework to include the postmerger phase would require a proper knowledge of the remnant object and its ringdown spectrum. Immediately following the merger, if the remnant features a ``clean" photon sphere, the prompt ringdown shows a striking similarity with that of a black-hole~\cite{cardoso:2017cqb}, although other features can arise when the remnant significantly deviates from BH description \cite{cardoso:2017cqb,Conklin:2019fcs}. Therefore, a complete inspiral-merger-ringdown (IMR) model should not only include $H$ as a parameter in the premerger regime, but also incorporate an appropriate ringdown model. In the present work, however, we do not introduce any non-BH deviations in the postmerger regime, and accordingly restrict the validity of our model to frequencies up to the BBH merger frequency. We truncate the complete waveform by tapering it from $f^{\rm peak}$ to $1.2f^{\rm peak}$.

The distinguishability of two frequency-domain waveforms $\tilde{h}_1(f)$ and $\tilde{h}_2(f)$ is defined by the match ($\mathcal{M}$) between them. One defines the inner product
\begin{equation}
    \braket{h_1}{h_2}=2\int_{f_{\rm low}}^{f_{\rm high}}\frac{\tilde{h}_1(f)\tilde{h}_2^\ast (f)+\tilde{h}_1^\ast (f)\tilde{h}_2(f)}{S_n(f)}\dd f\,,
\end{equation}
where $S_n(f)$ is the noise power spectral density. Match is defined by normalizing the inner product and maximizing over $t_c$ and $\phi_c$:
\begin{equation}
    \mathcal{M}=\underset{t_c,\phi_c}{\text{min}}\,\,\frac{\braket{h_1}{h_2}}{\sqrt{\braket{h_1}{h_1}\braket{h_2}{h_2}}}\,.
\end{equation}
By definition, $0\leqslant \mathcal{M}\leqslant 1$. Mismatch between these waveforms is the quantity $1-\mathcal{M}$.
In Fig.~\ref{fig:mismatch_H}, we show the mismatch between the full BBH waveforms and BCO waveforms with horizon parameter $H$ using the A+ design sensitivity curve of LIGO detector~\cite{Aasi:2013wya}, computed from 20~Hz up to merger. The white contours correspond to mismatches between the same waveforms, but computed only up to $f^{}_{\rm ISCO}=1/(6\sqrt{6}\pi M)$. 
The mismatches are shown as a function of the tidal heating parameter $H$ and the symmetric mass ratio $\eta$ for different total masses, $M=30,\,50,\,70\,$ and $90M_\odot$. The mismatch values decrease with increasing $H$, since by construction the waveforms are identical when $H=1$ (both BBHs), and differ the most when $H=0$ (one BBH, the other `point-particle'). Comparing with the white contours, we observe a substantial increase in the mismatch across the parameter space. For example, with $M=50M_\odot$, the colored points lying on the white contour of $10^{-5}\%$ report mismatches reaching values as high as $(1-\mathcal{M})\sim 0.007\%$. Overall, Fig.~\ref{fig:mismatch_H} demonstrates that the inclusion of the post-inspiral model of tidal heating significantly enhances the distinguishability of waveforms once the entire binary evolution is taken into account.

\section{Implications in BNS waveforms}\label{sec:bns}

\begin{figure}
    \centering
    \includegraphics[width=\linewidth]{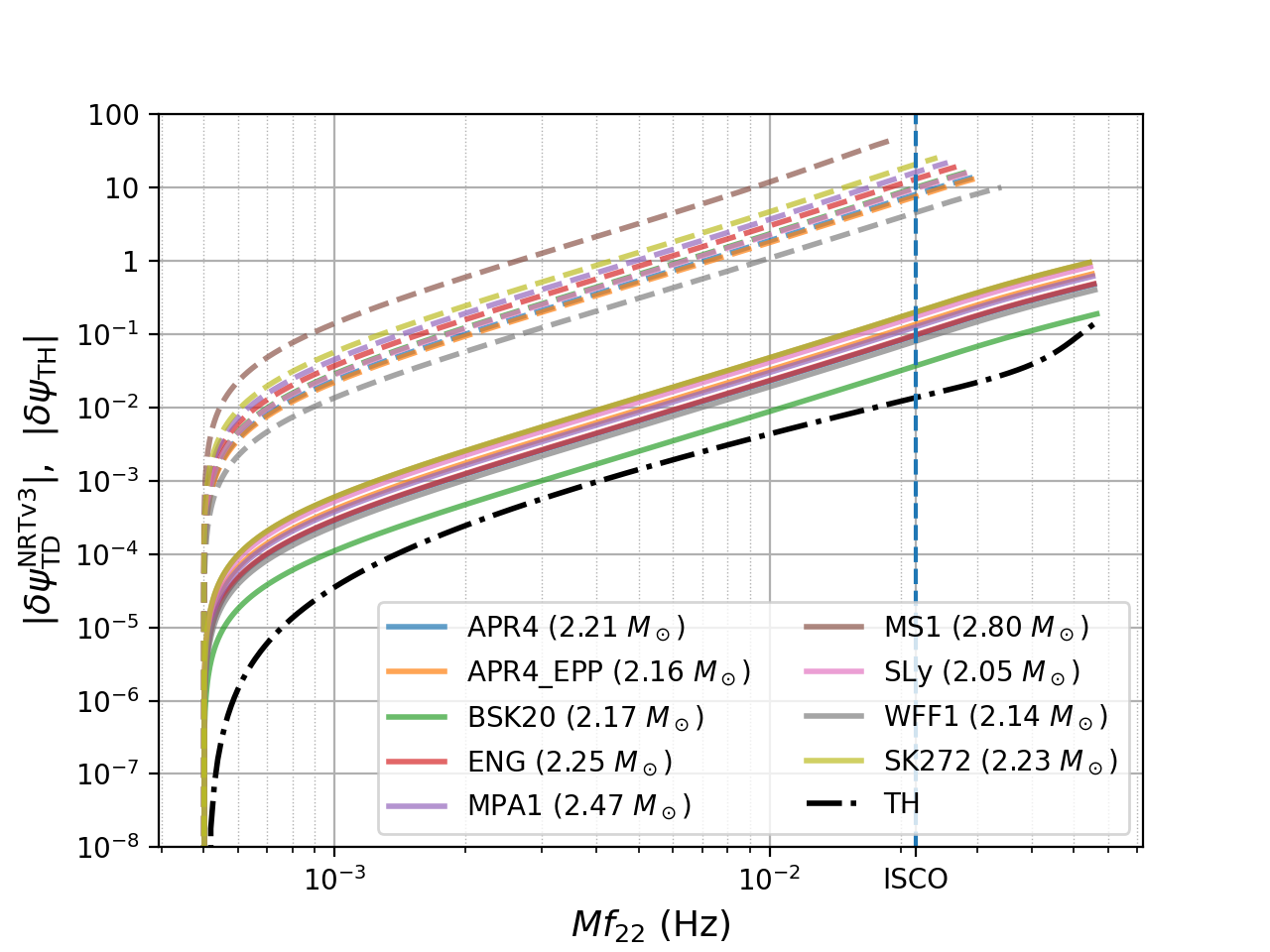}
    \caption{
    The colored solid lines show the total tidal contribution in phase (adiabatic and dynamical) for various NS EoSs with $m_{\rm max}+m_{\rm max}$  BNSs, $m_{\rm max}$ being the maximum allowed mass corresponding to each EoS (reported in parentheses). The dashed lines show the same with $1.4+1.4M_\odot$ systems. The dephasing is calculated using the \texttt{NRTidalv3} approximant~\cite{Abac:2023ujg}, up to the contact frequency corresponding to each configuration. The black dashed-dotted line shows the dephasing due to tidal heating in an equal-mass nonspinning binary black hole system, calculated up to the BBH merger frequency. All the phase curves are aligned by setting $\delta\psi=0$ at $Mf_{22}=0.0005$. The dimensionless ISCO frequency for a Schwarzschild black hole is also highlighted. 
    }
    \label{fig:dephasing_td_th}
\end{figure}

\begin{figure}
    \centering
    \includegraphics[width=\linewidth]{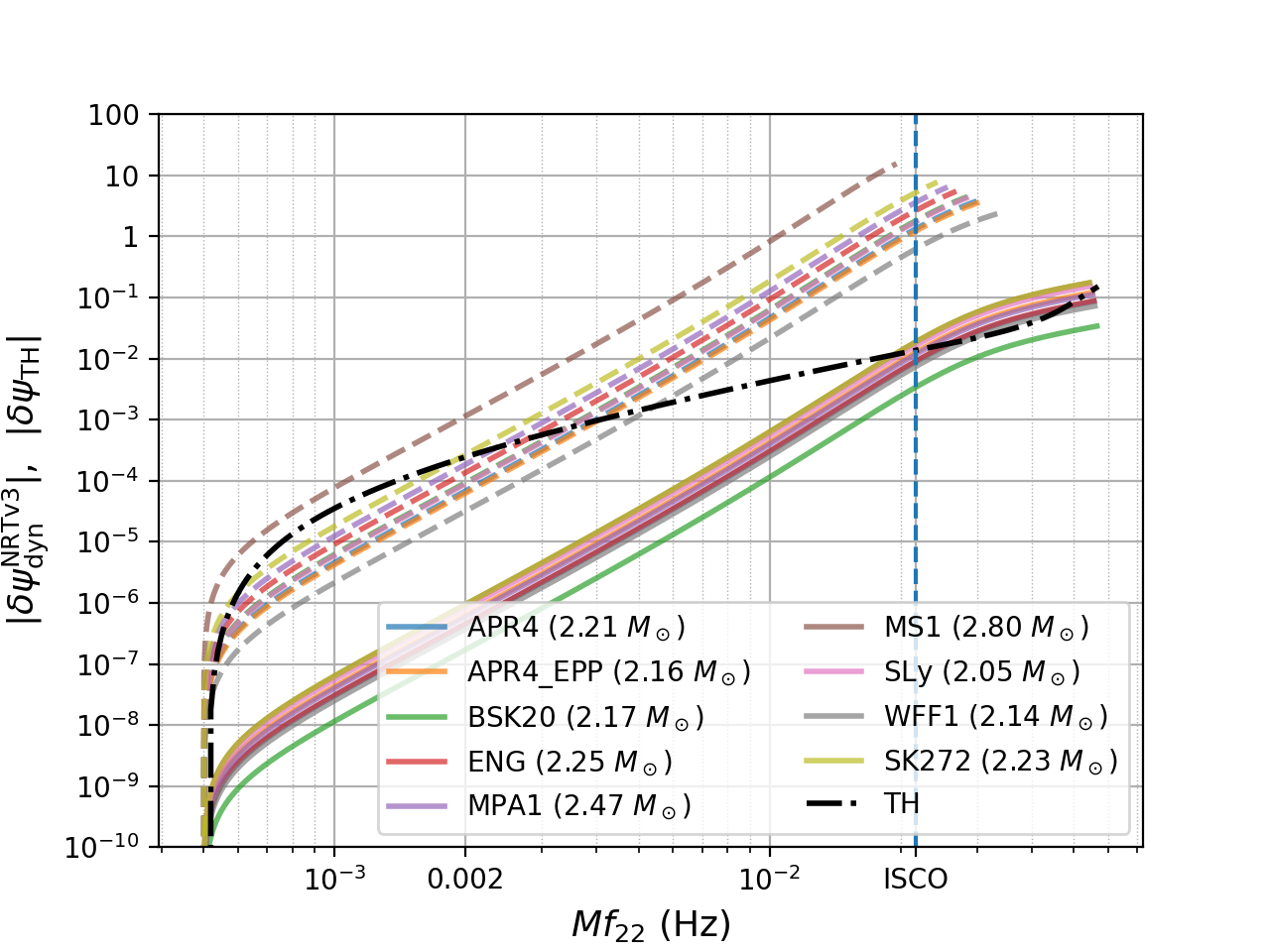}
    \caption{Same as Fig.~\ref{fig:dephasing_td_th}, but showing the dephasing in neutron stars solely arising from dynamical tides, as implemented in \texttt{NRTidalv3}. For $Mf\lesssim 0.002$, all the EoSs (except \texttt{MS1}) exhibit smaller dephasing than that from tidal heating in black holes for the dashed curves ($1.4+1.4M_\odot$ binaries). For the solid curves (heaviest possible NSs for each EoS), this behavior persists nearly up to the ISCO frequency.}
    \label{fig:dephasing_td_dyn_th}
\end{figure}

In modeling nonspinning BNS systems, the pre-merger phase difference between a BBH and a BNS waveform model is typically attributed entirely to the tidal deformability ($\Lambda$) of NSs~\cite{Dietrich:2019kaq,Dietrich:2020eud} as it is widely accepted that nonspinning BHs do not experience any conservative tidal deformation~\cite{Binnington:2009bb}. However, since NSs experience vanishing TH compared to BHs (see Refs.~\cite{Ghosh:2023vrx, Ghosh:2025glz,Saketh:2024juq,HegadeKR:2024slr} for an alternative scenario), not correcting for the effects of TH in the BBH system can introduce biases in the estimation of $\Lambda$ and NS fundamental modes excited by dynamical tides~\cite{Pratten:2019sed}. Figure~\ref{fig:dephasing_td_th} shows phase corrections in BNS gravitational waveforms due to their tidal deformability ($\delta\psi_{\rm TD}$)~\cite{Dietrich:2019kaq,Biswas:2021paf,lalsuite} for 9 different EoSs, and the phase correction in an equal-mass nonspinning binary black hole system due to horizon fluxes ($\delta\psi_{\rm TH}$). The phenomenological fits of $m(\Lambda)$ presented in Ref.~\cite{Biswas:2021paf} were inverted to estimate $\Lambda$ for a given NS mass corresponding to the different EoSs, and $\delta\psi_{\rm TD}$ was calculated from the \texttt{NRTidalv3}~\cite{Abac:2023ujg} approximant. For $1.4M_\odot+1.4M_\odot$ systems, differences of about three orders of magnitude are seen between $\delta\psi_{\rm TD}$ and $\delta\psi_{\rm TH}$. However, as the masses increase, $\delta\psi_{\rm TD}$ decreases concurrently with the tidal deformability. For the maximum masses allowed by the considered equations of state of neutron stars demonstrated in Fig.~\ref{fig:dephasing_td_th}, the TD and TH phases differ approximately by only one order of magnitude or lesser at higher frequencies. 
These estimates underscore the importance of 
accounting for this aspect in BNS modeling, especially for heavier systems.

Total contribution from the tidal deformation of NSs in a BNS also includes the contribution from dynamical tides, the corrections due to which are expected to be small compared to that from adiabatic tides. The tidal approximant \texttt{NRTidalv3} includes this effect in its total tidal phase. To demonstrate how $\delta\psi_{\rm TH}$ compares with this effect, we plot the phase correction $\delta \psi_{\rm dyn}^{\rm NRTv3}$ from dynamical tides  alongside $\delta\psi_{\rm TH}$ in Fig.~\ref{fig:dephasing_td_dyn_th}. We compute $\delta \psi_{\rm dyn}^{\rm NRTv3}$ by first evaluating the total dephasing $\delta\psi_{\rm TD}^{\rm NRTv3}$, and then setting the amplification factor $\bar{k}_2^{\rm eff}(\hat{\omega})=1$ (see Eq. (24) of Ref~\cite{Abac:2023ujg}) to estimate it \emph{without} the dynamical tides part. Then, 
\begin{equation}
   \delta \psi_{\rm dyn}^{\rm NRTv3}=\delta\psi_{\rm TD}^{\rm NRTv3}-\delta\psi^{\rm NRTv3}_{\rm TD}\big|_{\bar{k}_2^{\rm eff}(\hat{\omega})=1}\,. 
\end{equation}
For $1.4+1.4M_\odot$ BNSs, the dephasing from tidal heating in black holes surpasses that from dynamical tides at lower frequencies -- up to $Mf\sim 0.002$ -- for almost all of the EoSs considered here (with the only exception of \texttt{MS1}). For a total mass $M=2.8M_\odot$, this corresponds to $f\sim 145$ Hz. For heavier BNS systems the relative importance of TH rises further. These two effects are, therefore, of comparable significance in these regimes. In future GW detectors, phase modulation due to dynamical tides will help constrain the $f$-mode frequencies of NSs to within a few tens of Hz~\cite{Pratten:2019sed}. Figure~\ref{fig:dephasing_td_dyn_th} suggests that removing the systematics due to TH of BHs will be essential in such analyses in the third-generation GW detectors.

\section{Conclusion}\label{sec:conclusion}

The primary objective of this work is to highlight the key features of the mass evolution of black holes in a binary systems as reported from numerical relativity simulations and to introduce a framework to model its impact on the binary's inspiral rate. We describe the construction of  such a model in the frequency-domain, based on the horizon data provided by the SXS catalog.
In modeling gravitational waveforms from binaries of generic compact objects, this approach may have important implications. Firstly, We show that the analytical calculations from the post-Newtonian approximation considerably underestimate the horizon fluxes as predicted by numerical relativity data, by up to approximately three orders of magnitude close to merger (Fig.~\ref{fig:flux_q10}), for nonspinning black hole binaries within the mass-ratio range $1\leqslant q\leqslant 10$.
By leveraging the substantial amount of flux lost into the horizons in late stages of the coalescence, the present model provides a more robust tool to identify black-hole mimickers than the inspiral-only waveforms equipped with analytical expressions. Secondly, a model of the horizon fluxes enables one to construct refined point-particle waveforms free from all the finite-size effects of its components. 
With the third-generation detectors on the horizon~\cite{Evans:2021gyd, Abac:2025saz}, minimizing systematic errors in gravitational waveform modeling is crucial for testing the black hole nature of the observed compact objects, and probing subtle matter effects in neutron stars. A central contribution of this work is 
to demonstrate, using numerical relativity data, the extent to which a black hole binary evolution deviates from a corresponding `point-particle' binary due to the presence of horizons, and to develop a model that captures this difference. To avoid introducing systematics in compact binary waveform models for non-black hole objects, it is essential to account for the imprints of this phenomenon in binary black hole systems, and optimize the waveform baselines accordingly.

\begin{acknowledgments}
We are grateful to Nathan K. Johnson-McDaniel for carefully reading the manuscript and providing insightful comments and suggestions. SM thanks Harald Pfeiffer for his valuable feedback on an earlier version of the manuscript, and P. Ajith, Prayush Kumar for their inputs. 
SD thanks Francisco Jiménez Forteza and Andrea Maselli for helpful discussions. We also appreciate the anonymous referees for their constructive suggestions, which helped improve the manuscript.
SD acknowledges financial support from MUR, PNRR - Missione 4 - Componente 2 - Investimento 1.2 - finanziato dall'Unione europea - NextGenerationEU (cod. id.: SOE2024\_0000167, CUP:D13C25000660001). KSP acknowledges support from STFC grant ST/V005677/1.
The authors are grateful for computational resources provided by the LIGO Laboratory (CIT) and supported by
the National Science Foundation Grants PHY-0757058
and PHY-0823459 as well as the Sarathi cluster at the Inter-University Center for Astronomy \& Astrophysics, Pune, India. We also acknowledge the NSF Grant PHY-2309352. This study used software packages Numpy~\cite{Harris:2020xlr}, Matplotlib~\cite{Hunter:2007ouj}, Scipy~\cite{2020SciPy-NMeth} and the Simulating eXtreme Spacetimes package~\cite{boyle_2024_13714441}. This manuscript has the LIGO document number P2500371. S.M. acknowledges support of the Department of Atomic Energy, Government of India, under Project Nos. RTI4019 and RTI4013.
\end{acknowledgments}

%


\begin{thebibliography}{70}%
\makeatletter
\providecommand \@ifxundefined [1]{%
 \@ifx{#1\undefined}
}%
\providecommand \@ifnum [1]{%
 \ifnum #1\expandafter \@firstoftwo
 \else \expandafter \@secondoftwo
 \fi
}%
\providecommand \@ifx [1]{%
 \ifx #1\expandafter \@firstoftwo
 \else \expandafter \@secondoftwo
 \fi
}%
\providecommand \natexlab [1]{#1}%
\providecommand \enquote  [1]{``#1''}%
\providecommand \bibnamefont  [1]{#1}%
\providecommand \bibfnamefont [1]{#1}%
\providecommand \citenamefont [1]{#1}%
\providecommand \href@noop [0]{\@secondoftwo}%
\providecommand \href [0]{\begingroup \@sanitize@url \@href}%
\providecommand \@href[1]{\@@startlink{#1}\@@href}%
\providecommand \@@href[1]{\endgroup#1\@@endlink}%
\providecommand \@sanitize@url [0]{\catcode `\\12\catcode `\$12\catcode `\&12\catcode `\#12\catcode `\^12\catcode `\_12\catcode `\%12\relax}%
\providecommand \@@startlink[1]{}%
\providecommand \@@endlink[0]{}%
\providecommand \url  [0]{\begingroup\@sanitize@url \@url }%
\providecommand \@url [1]{\endgroup\@href {#1}{\urlprefix }}%
\providecommand \urlprefix  [0]{URL }%
\providecommand \Eprint [0]{\href }%
\providecommand \doibase [0]{https://doi.org/}%
\providecommand \selectlanguage [0]{\@gobble}%
\providecommand \bibinfo  [0]{\@secondoftwo}%
\providecommand \bibfield  [0]{\@secondoftwo}%
\providecommand \translation [1]{[#1]}%
\providecommand \BibitemOpen [0]{}%
\providecommand \bibitemStop [0]{}%
\providecommand \bibitemNoStop [0]{.\EOS\space}%
\providecommand \EOS [0]{\spacefactor3000\relax}%
\providecommand \BibitemShut  [1]{\csname bibitem#1\endcsname}%
\let\auto@bib@innerbib\@empty
\bibitem [{\citenamefont {Hartle}(1973)}]{Hartle:1973zz}%
  \BibitemOpen
  \bibfield  {author} {\bibinfo {author} {\bibfnamefont {J.~B.}\ \bibnamefont {Hartle}},\ }\bibfield  {title} {\bibinfo {title} {{Tidal Friction in Slowly Rotating Black Holes}},\ }\href {https://doi.org/10.1103/PhysRevD.8.1010} {\bibfield  {journal} {\bibinfo  {journal} {Phys. Rev. D}\ }\textbf {\bibinfo {volume} {8}},\ \bibinfo {pages} {1010} (\bibinfo {year} {1973})}\BibitemShut {NoStop}%
\bibitem [{\citenamefont {{Vega}}\ \emph {et~al.}(2011)\citenamefont {{Vega}}, \citenamefont {{Poisson}},\ and\ \citenamefont {{Massey}}}]{Vega2011}%
  \BibitemOpen
  \bibfield  {author} {\bibinfo {author} {\bibfnamefont {I.}~\bibnamefont {{Vega}}}, \bibinfo {author} {\bibfnamefont {E.}~\bibnamefont {{Poisson}}},\ and\ \bibinfo {author} {\bibfnamefont {R.}~\bibnamefont {{Massey}}},\ }\bibfield  {title} {\bibinfo {title} {{Intrinsic and extrinsic geometries of a tidally deformed black hole}},\ }\href {https://doi.org/10.1088/0264-9381/28/17/175006} {\bibfield  {journal} {\bibinfo  {journal} {Classical and Quantum Gravity}\ }\textbf {\bibinfo {volume} {28}},\ \bibinfo {eid} {175006} (\bibinfo {year} {2011})},\ \Eprint {https://arxiv.org/abs/1106.0510} {arXiv:1106.0510 [gr-qc]} \BibitemShut {NoStop}%
\bibitem [{\citenamefont {O'Sullivan}\ and\ \citenamefont {Hughes}(2014)}]{OSullivan:2014ywd}%
  \BibitemOpen
  \bibfield  {author} {\bibinfo {author} {\bibfnamefont {S.}~\bibnamefont {O'Sullivan}}\ and\ \bibinfo {author} {\bibfnamefont {S.~A.}\ \bibnamefont {Hughes}},\ }\bibfield  {title} {\bibinfo {title} {{Strong-field tidal distortions of rotating black holes: Formalism and results for circular, equatorial orbits}},\ }\href {https://doi.org/10.1103/PhysRevD.91.109901} {\bibfield  {journal} {\bibinfo  {journal} {Phys. Rev. D}\ }\textbf {\bibinfo {volume} {90}},\ \bibinfo {pages} {124039} (\bibinfo {year} {2014})},\ \bibinfo {note} {[Erratum: Phys.Rev.D 91, 109901 (2015)]},\ \Eprint {https://arxiv.org/abs/1407.6983} {arXiv:1407.6983 [gr-qc]} \BibitemShut {NoStop}%
\bibitem [{\citenamefont {O'Sullivan}\ and\ \citenamefont {Hughes}(2016)}]{OSullivan:2015lni}%
  \BibitemOpen
  \bibfield  {author} {\bibinfo {author} {\bibfnamefont {S.}~\bibnamefont {O'Sullivan}}\ and\ \bibinfo {author} {\bibfnamefont {S.~A.}\ \bibnamefont {Hughes}},\ }\bibfield  {title} {\bibinfo {title} {{Strong-field tidal distortions of rotating black holes: II. Horizon dynamics from eccentric and inclined orbits}},\ }\href {https://doi.org/10.1103/PhysRevD.94.044057} {\bibfield  {journal} {\bibinfo  {journal} {Phys. Rev. D}\ }\textbf {\bibinfo {volume} {94}},\ \bibinfo {pages} {044057} (\bibinfo {year} {2016})},\ \Eprint {https://arxiv.org/abs/1505.03809} {arXiv:1505.03809 [gr-qc]} \BibitemShut {NoStop}%
\bibitem [{\citenamefont {Prasad}\ \emph {et~al.}(2022)\citenamefont {Prasad}, \citenamefont {Gupta}, \citenamefont {Bose},\ and\ \citenamefont {Krishnan}}]{Prasad:2021dfr}%
  \BibitemOpen
  \bibfield  {author} {\bibinfo {author} {\bibfnamefont {V.}~\bibnamefont {Prasad}}, \bibinfo {author} {\bibfnamefont {A.}~\bibnamefont {Gupta}}, \bibinfo {author} {\bibfnamefont {S.}~\bibnamefont {Bose}},\ and\ \bibinfo {author} {\bibfnamefont {B.}~\bibnamefont {Krishnan}},\ }\bibfield  {title} {\bibinfo {title} {{Tidal deformation of dynamical horizons in binary black hole mergers}},\ }\href {https://doi.org/10.1103/PhysRevD.105.044019} {\bibfield  {journal} {\bibinfo  {journal} {Phys. Rev. D}\ }\textbf {\bibinfo {volume} {105}},\ \bibinfo {pages} {044019} (\bibinfo {year} {2022})},\ \Eprint {https://arxiv.org/abs/2106.02595} {arXiv:2106.02595 [gr-qc]} \BibitemShut {NoStop}%
\bibitem [{\citenamefont {Datta}\ \emph {et~al.}(2021)\citenamefont {Datta}, \citenamefont {Phukon},\ and\ \citenamefont {Bose}}]{datta2020recognizing}%
  \BibitemOpen
  \bibfield  {author} {\bibinfo {author} {\bibfnamefont {S.}~\bibnamefont {Datta}}, \bibinfo {author} {\bibfnamefont {K.~S.}\ \bibnamefont {Phukon}},\ and\ \bibinfo {author} {\bibfnamefont {S.}~\bibnamefont {Bose}},\ }\bibfield  {title} {\bibinfo {title} {{Recognizing black holes in gravitational-wave observations: Challenges in telling apart impostors in mass-gap binaries}},\ }\href {https://doi.org/10.1103/PhysRevD.104.084006} {\bibfield  {journal} {\bibinfo  {journal} {Phys. Rev. D}\ }\textbf {\bibinfo {volume} {104}},\ \bibinfo {pages} {084006} (\bibinfo {year} {2021})},\ \Eprint {https://arxiv.org/abs/2004.05974} {arXiv:2004.05974 [gr-qc]} \BibitemShut {NoStop}%
\bibitem [{\citenamefont {Mukherjee}\ \emph {et~al.}(2022)\citenamefont {Mukherjee}, \citenamefont {Datta}, \citenamefont {Tiwari}, \citenamefont {Phukon},\ and\ \citenamefont {Bose}}]{Mukherjee:2022wws}%
  \BibitemOpen
  \bibfield  {author} {\bibinfo {author} {\bibfnamefont {S.}~\bibnamefont {Mukherjee}}, \bibinfo {author} {\bibfnamefont {S.}~\bibnamefont {Datta}}, \bibinfo {author} {\bibfnamefont {S.}~\bibnamefont {Tiwari}}, \bibinfo {author} {\bibfnamefont {K.~S.}\ \bibnamefont {Phukon}},\ and\ \bibinfo {author} {\bibfnamefont {S.}~\bibnamefont {Bose}},\ }\bibfield  {title} {\bibinfo {title} {{Toward establishing the presence or absence of horizons in coalescing binaries of compact objects by using their gravitational wave signals}},\ }\href {https://doi.org/10.1103/PhysRevD.106.104032} {\bibfield  {journal} {\bibinfo  {journal} {Phys. Rev. D}\ }\textbf {\bibinfo {volume} {106}},\ \bibinfo {pages} {104032} (\bibinfo {year} {2022})},\ \Eprint {https://arxiv.org/abs/2202.08661} {arXiv:2202.08661 [gr-qc]} \BibitemShut {NoStop}%
\bibitem [{\citenamefont {{Hawking}}\ and\ \citenamefont {{Hartle}}(1972)}]{HawkingHartle}%
  \BibitemOpen
  \bibfield  {author} {\bibinfo {author} {\bibfnamefont {S.~W.}\ \bibnamefont {{Hawking}}}\ and\ \bibinfo {author} {\bibfnamefont {J.~B.}\ \bibnamefont {{Hartle}}},\ }\bibfield  {title} {\bibinfo {title} {{Energy and angular momentum flow into a black hole}},\ }\href {https://doi.org/10.1007/BF01645515} {\bibfield  {journal} {\bibinfo  {journal} {Communications in Mathematical Physics}\ }\textbf {\bibinfo {volume} {27}},\ \bibinfo {pages} {283} (\bibinfo {year} {1972})}\BibitemShut {NoStop}%
\bibitem [{\citenamefont {Hughes}(2001)}]{Hughes:2001jr}%
  \BibitemOpen
  \bibfield  {author} {\bibinfo {author} {\bibfnamefont {S.~A.}\ \bibnamefont {Hughes}},\ }\bibfield  {title} {\bibinfo {title} {{Evolution of circular, nonequatorial orbits of Kerr black holes due to gravitational wave emission. II. Inspiral trajectories and gravitational wave forms}},\ }\href {https://doi.org/10.1103/PhysRevD.64.064004} {\bibfield  {journal} {\bibinfo  {journal} {Phys. Rev. D}\ }\textbf {\bibinfo {volume} {64}},\ \bibinfo {pages} {064004} (\bibinfo {year} {2001})},\ \bibinfo {note} {[Erratum: Phys.Rev.D 88, 109902 (2013)]},\ \Eprint {https://arxiv.org/abs/gr-qc/0104041} {arXiv:gr-qc/0104041} \BibitemShut {NoStop}%
\bibitem [{\citenamefont {Taracchini}\ \emph {et~al.}(2013)\citenamefont {Taracchini}, \citenamefont {Buonanno}, \citenamefont {Hughes},\ and\ \citenamefont {Khanna}}]{Taracchini:2013wfa}%
  \BibitemOpen
  \bibfield  {author} {\bibinfo {author} {\bibfnamefont {A.}~\bibnamefont {Taracchini}}, \bibinfo {author} {\bibfnamefont {A.}~\bibnamefont {Buonanno}}, \bibinfo {author} {\bibfnamefont {S.~A.}\ \bibnamefont {Hughes}},\ and\ \bibinfo {author} {\bibfnamefont {G.}~\bibnamefont {Khanna}},\ }\bibfield  {title} {\bibinfo {title} {{Modeling the horizon-absorbed gravitational flux for equatorial-circular orbits in Kerr spacetime}},\ }\href {https://doi.org/10.1103/PhysRevD.88.044001} {\bibfield  {journal} {\bibinfo  {journal} {Phys. Rev. D}\ }\textbf {\bibinfo {volume} {88}},\ \bibinfo {pages} {044001} (\bibinfo {year} {2013})},\ \bibinfo {note} {[Erratum: Phys.Rev.D 88, 109903 (2013)]},\ \Eprint {https://arxiv.org/abs/1305.2184} {arXiv:1305.2184 [gr-qc]} \BibitemShut {NoStop}%
\bibitem [{\citenamefont {Harms}\ \emph {et~al.}(2014)\citenamefont {Harms}, \citenamefont {Bernuzzi}, \citenamefont {Nagar},\ and\ \citenamefont {Zenginoglu}}]{Harms:2014dqa}%
  \BibitemOpen
  \bibfield  {author} {\bibinfo {author} {\bibfnamefont {E.}~\bibnamefont {Harms}}, \bibinfo {author} {\bibfnamefont {S.}~\bibnamefont {Bernuzzi}}, \bibinfo {author} {\bibfnamefont {A.}~\bibnamefont {Nagar}},\ and\ \bibinfo {author} {\bibfnamefont {A.}~\bibnamefont {Zenginoglu}},\ }\bibfield  {title} {\bibinfo {title} {{A new gravitational wave generation algorithm for particle perturbations of the Kerr spacetime}},\ }\href {https://doi.org/10.1088/0264-9381/31/24/245004} {\bibfield  {journal} {\bibinfo  {journal} {Class. Quant. Grav.}\ }\textbf {\bibinfo {volume} {31}},\ \bibinfo {pages} {245004} (\bibinfo {year} {2014})},\ \Eprint {https://arxiv.org/abs/1406.5983} {arXiv:1406.5983 [gr-qc]} \BibitemShut {NoStop}%
\bibitem [{\citenamefont {Datta}\ and\ \citenamefont {Bose}(2019)}]{Datta:2019euh}%
  \BibitemOpen
  \bibfield  {author} {\bibinfo {author} {\bibfnamefont {S.}~\bibnamefont {Datta}}\ and\ \bibinfo {author} {\bibfnamefont {S.}~\bibnamefont {Bose}},\ }\bibfield  {title} {\bibinfo {title} {{Probing the nature of central objects in extreme-mass-ratio inspirals with gravitational waves}},\ }\href {https://doi.org/10.1103/PhysRevD.99.084001} {\bibfield  {journal} {\bibinfo  {journal} {Phys. Rev.}\ }\textbf {\bibinfo {volume} {D99}},\ \bibinfo {pages} {084001} (\bibinfo {year} {2019})},\ \Eprint {https://arxiv.org/abs/1902.01723} {arXiv:1902.01723 [gr-qc]} \BibitemShut {NoStop}%
\bibitem [{\citenamefont {Datta}\ \emph {et~al.}(2020)\citenamefont {Datta}, \citenamefont {Brito}, \citenamefont {Bose}, \citenamefont {Pani},\ and\ \citenamefont {Hughes}}]{Datta:2019epe}%
  \BibitemOpen
  \bibfield  {author} {\bibinfo {author} {\bibfnamefont {S.}~\bibnamefont {Datta}}, \bibinfo {author} {\bibfnamefont {R.}~\bibnamefont {Brito}}, \bibinfo {author} {\bibfnamefont {S.}~\bibnamefont {Bose}}, \bibinfo {author} {\bibfnamefont {P.}~\bibnamefont {Pani}},\ and\ \bibinfo {author} {\bibfnamefont {S.~A.}\ \bibnamefont {Hughes}},\ }\bibfield  {title} {\bibinfo {title} {{Tidal heating as a discriminator for horizons in extreme mass ratio inspirals}},\ }\href {https://doi.org/10.1103/PhysRevD.101.044004} {\bibfield  {journal} {\bibinfo  {journal} {Phys. Rev.}\ }\textbf {\bibinfo {volume} {D101}},\ \bibinfo {pages} {044004} (\bibinfo {year} {2020})},\ \Eprint {https://arxiv.org/abs/1910.07841} {arXiv:1910.07841 [gr-qc]} \BibitemShut {NoStop}%
\bibitem [{\citenamefont {Datta}\ \emph {et~al.}(2024)\citenamefont {Datta}, \citenamefont {Brito}, \citenamefont {Hughes}, \citenamefont {Klinger},\ and\ \citenamefont {Pani}}]{Datta:2024vll}%
  \BibitemOpen
  \bibfield  {author} {\bibinfo {author} {\bibfnamefont {S.}~\bibnamefont {Datta}}, \bibinfo {author} {\bibfnamefont {R.}~\bibnamefont {Brito}}, \bibinfo {author} {\bibfnamefont {S.~A.}\ \bibnamefont {Hughes}}, \bibinfo {author} {\bibfnamefont {T.}~\bibnamefont {Klinger}},\ and\ \bibinfo {author} {\bibfnamefont {P.}~\bibnamefont {Pani}},\ }\bibfield  {title} {\bibinfo {title} {{Tidal heating as a discriminator for horizons in equatorial eccentric extreme mass ratio inspirals}},\ }\href {https://doi.org/10.1103/PhysRevD.110.024048} {\bibfield  {journal} {\bibinfo  {journal} {Phys. Rev. D}\ }\textbf {\bibinfo {volume} {110}},\ \bibinfo {pages} {024048} (\bibinfo {year} {2024})},\ \Eprint {https://arxiv.org/abs/2404.04013} {arXiv:2404.04013 [gr-qc]} \BibitemShut {NoStop}%
\bibitem [{\citenamefont {Colpi}\ \emph {et~al.}(2024)\citenamefont {Colpi} \emph {et~al.}}]{Colpi:2024xhw}%
  \BibitemOpen
  \bibfield  {author} {\bibinfo {author} {\bibfnamefont {M.}~\bibnamefont {Colpi}} \emph {et~al.},\ }\bibfield  {title} {\bibinfo {title} {{LISA Definition Study Report}},\ }\href@noop {} {\  (\bibinfo {year} {2024})},\ \Eprint {https://arxiv.org/abs/2402.07571} {arXiv:2402.07571 [astro-ph.CO]} \BibitemShut {NoStop}%
\bibitem [{\citenamefont {Aasi}\ \emph {et~al.}(2015)\citenamefont {Aasi} \emph {et~al.}}]{TheLIGOScientific:2014jea}%
  \BibitemOpen
  \bibfield  {author} {\bibinfo {author} {\bibfnamefont {J.}~\bibnamefont {Aasi}} \emph {et~al.} (\bibinfo {collaboration} {LIGO Scientific}),\ }\bibfield  {title} {\bibinfo {title} {{Advanced LIGO}},\ }\href {https://doi.org/10.1088/0264-9381/32/7/074001} {\bibfield  {journal} {\bibinfo  {journal} {Class. Quant. Grav.}\ }\textbf {\bibinfo {volume} {32}},\ \bibinfo {pages} {074001} (\bibinfo {year} {2015})},\ \Eprint {https://arxiv.org/abs/1411.4547} {arXiv:1411.4547 [gr-qc]} \BibitemShut {NoStop}%
\bibitem [{\citenamefont {Acernese}\ \emph {et~al.}(2015)\citenamefont {Acernese} \emph {et~al.}}]{VIRGO:2014yos}%
  \BibitemOpen
  \bibfield  {author} {\bibinfo {author} {\bibfnamefont {F.}~\bibnamefont {Acernese}} \emph {et~al.} (\bibinfo {collaboration} {VIRGO}),\ }\bibfield  {title} {\bibinfo {title} {{Advanced Virgo: a second-generation interferometric gravitational wave detector}},\ }\href {https://doi.org/10.1088/0264-9381/32/2/024001} {\bibfield  {journal} {\bibinfo  {journal} {Class. Quant. Grav.}\ }\textbf {\bibinfo {volume} {32}},\ \bibinfo {pages} {024001} (\bibinfo {year} {2015})},\ \Eprint {https://arxiv.org/abs/1408.3978} {arXiv:1408.3978 [gr-qc]} \BibitemShut {NoStop}%
\bibitem [{\citenamefont {Abbott}\ \emph {et~al.}(2020{\natexlab{a}})\citenamefont {Abbott} \emph {et~al.}}]{Aasi:2013wya}%
  \BibitemOpen
  \bibfield  {author} {\bibinfo {author} {\bibfnamefont {B.~P.}\ \bibnamefont {Abbott}} \emph {et~al.} (\bibinfo {collaboration} {KAGRA Collaboration, LIGO Scientific Collaboration, and Virgo Collaboration}),\ }\bibfield  {title} {\bibinfo {title} {{Prospects for Observing and Localizing Gravitational-Wave Transients with Advanced LIGO, Advanced Virgo and KAGRA}},\ }\href {https://doi.org/10.1007/s41114-020-00026-9} {\bibfield  {journal} {\bibinfo  {journal} {Living Rev. Relativity}\ }\textbf {\bibinfo {volume} {23}},\ \bibinfo {pages} {3} (\bibinfo {year} {2020}{\natexlab{a}})},\ \bibinfo {note} {noise curves available from \url{https://dcc.ligo.org/LIGO-T2000012/public}},\ \Eprint {https://arxiv.org/abs/1304.0670} {arXiv:1304.0670 [gr-qc]} \BibitemShut {NoStop}%
\bibitem [{\citenamefont {Reitze}\ \emph {et~al.}(2019)\citenamefont {Reitze} \emph {et~al.}}]{Reitze:2019iox}%
  \BibitemOpen
  \bibfield  {author} {\bibinfo {author} {\bibfnamefont {D.}~\bibnamefont {Reitze}} \emph {et~al.},\ }\bibfield  {title} {\bibinfo {title} {{Cosmic Explorer: The U.S. Contribution to Gravitational-Wave Astronomy beyond LIGO}},\ }\href@noop {} {\bibfield  {journal} {\bibinfo  {journal} {Bull. Am. Astron. Soc.}\ }\textbf {\bibinfo {volume} {51}},\ \bibinfo {pages} {035} (\bibinfo {year} {2019})},\ \Eprint {https://arxiv.org/abs/1907.04833} {arXiv:1907.04833 [astro-ph.IM]} \BibitemShut {NoStop}%
\bibitem [{\citenamefont {Abac}\ \emph {et~al.}(2025)\citenamefont {Abac} \emph {et~al.}}]{Abac:2025saz}%
  \BibitemOpen
  \bibfield  {author} {\bibinfo {author} {\bibfnamefont {A.}~\bibnamefont {Abac}} \emph {et~al.},\ }\bibfield  {title} {\bibinfo {title} {{The Science of the Einstein Telescope}},\ }\href@noop {} {\  (\bibinfo {year} {2025})},\ \Eprint {https://arxiv.org/abs/2503.12263} {arXiv:2503.12263 [gr-qc]} \BibitemShut {NoStop}%
\bibitem [{\citenamefont {Cardoso}\ and\ \citenamefont {Pani}(2019)}]{Cardoso:2019rvt}%
  \BibitemOpen
  \bibfield  {author} {\bibinfo {author} {\bibfnamefont {V.}~\bibnamefont {Cardoso}}\ and\ \bibinfo {author} {\bibfnamefont {P.}~\bibnamefont {Pani}},\ }\bibfield  {title} {\bibinfo {title} {{Testing the nature of dark compact objects: a status report}},\ }\href {https://doi.org/10.1007/s41114-019-0020-4} {\bibfield  {journal} {\bibinfo  {journal} {Living Rev. Rel.}\ }\textbf {\bibinfo {volume} {22}},\ \bibinfo {pages} {4} (\bibinfo {year} {2019})},\ \Eprint {https://arxiv.org/abs/1904.05363} {arXiv:1904.05363 [gr-qc]} \BibitemShut {NoStop}%
\bibitem [{\citenamefont {Maselli}\ \emph {et~al.}(2018)\citenamefont {Maselli}, \citenamefont {Pani}, \citenamefont {Cardoso}, \citenamefont {Abdelsalhin}, \citenamefont {Gualtieri},\ and\ \citenamefont {Ferrari}}]{Maselli:2017cmm}%
  \BibitemOpen
  \bibfield  {author} {\bibinfo {author} {\bibfnamefont {A.}~\bibnamefont {Maselli}}, \bibinfo {author} {\bibfnamefont {P.}~\bibnamefont {Pani}}, \bibinfo {author} {\bibfnamefont {V.}~\bibnamefont {Cardoso}}, \bibinfo {author} {\bibfnamefont {T.}~\bibnamefont {Abdelsalhin}}, \bibinfo {author} {\bibfnamefont {L.}~\bibnamefont {Gualtieri}},\ and\ \bibinfo {author} {\bibfnamefont {V.}~\bibnamefont {Ferrari}},\ }\bibfield  {title} {\bibinfo {title} {{Probing Planckian corrections at the horizon scale with LISA binaries}},\ }\href {https://doi.org/10.1103/PhysRevLett.120.081101} {\bibfield  {journal} {\bibinfo  {journal} {Phys. Rev. Lett.}\ }\textbf {\bibinfo {volume} {120}},\ \bibinfo {pages} {081101} (\bibinfo {year} {2018})},\ \Eprint {https://arxiv.org/abs/1703.10612} {arXiv:1703.10612 [gr-qc]} \BibitemShut {NoStop}%
\bibitem [{\citenamefont {Datta}(2020)}]{Datta:2020rvo}%
  \BibitemOpen
  \bibfield  {author} {\bibinfo {author} {\bibfnamefont {S.}~\bibnamefont {Datta}},\ }\bibfield  {title} {\bibinfo {title} {{Tidal heating of Quantum Black Holes and their imprints on gravitational waves}},\ }\href {https://doi.org/10.1103/PhysRevD.102.064040} {\bibfield  {journal} {\bibinfo  {journal} {Phys. Rev. D}\ }\textbf {\bibinfo {volume} {102}},\ \bibinfo {pages} {064040} (\bibinfo {year} {2020})},\ \Eprint {https://arxiv.org/abs/2002.04480} {arXiv:2002.04480 [gr-qc]} \BibitemShut {NoStop}%
\bibitem [{\citenamefont {Maggio}\ \emph {et~al.}(2021)\citenamefont {Maggio}, \citenamefont {van~de Meent},\ and\ \citenamefont {Pani}}]{Maggio:2021uge}%
  \BibitemOpen
  \bibfield  {author} {\bibinfo {author} {\bibfnamefont {E.}~\bibnamefont {Maggio}}, \bibinfo {author} {\bibfnamefont {M.}~\bibnamefont {van~de Meent}},\ and\ \bibinfo {author} {\bibfnamefont {P.}~\bibnamefont {Pani}},\ }\bibfield  {title} {\bibinfo {title} {{Extreme mass-ratio inspirals around a spinning horizonless compact object}},\ }\href {https://doi.org/10.1103/PhysRevD.104.104026} {\bibfield  {journal} {\bibinfo  {journal} {Phys. Rev. D}\ }\textbf {\bibinfo {volume} {104}},\ \bibinfo {pages} {104026} (\bibinfo {year} {2021})},\ \Eprint {https://arxiv.org/abs/2106.07195} {arXiv:2106.07195 [gr-qc]} \BibitemShut {NoStop}%
\bibitem [{\citenamefont {Glampedakis}\ \emph {et~al.}(2014)\citenamefont {Glampedakis}, \citenamefont {Kapadia},\ and\ \citenamefont {Kennefick}}]{Glampedakis:2013jya}%
  \BibitemOpen
  \bibfield  {author} {\bibinfo {author} {\bibfnamefont {K.}~\bibnamefont {Glampedakis}}, \bibinfo {author} {\bibfnamefont {S.~J.}\ \bibnamefont {Kapadia}},\ and\ \bibinfo {author} {\bibfnamefont {D.}~\bibnamefont {Kennefick}},\ }\bibfield  {title} {\bibinfo {title} {{Superradiance-tidal friction correspondence}},\ }\href {https://doi.org/10.1103/PhysRevD.89.024007} {\bibfield  {journal} {\bibinfo  {journal} {Phys. Rev.}\ }\textbf {\bibinfo {volume} {D89}},\ \bibinfo {pages} {024007} (\bibinfo {year} {2014})},\ \Eprint {https://arxiv.org/abs/1312.1912} {arXiv:1312.1912 [gr-qc]} \BibitemShut {NoStop}%
\bibitem [{\citenamefont {Rhoades}\ and\ \citenamefont {Ruffini}(1974)}]{Rhoades:1974fn}%
  \BibitemOpen
  \bibfield  {author} {\bibinfo {author} {\bibfnamefont {C.~E.}\ \bibnamefont {Rhoades}, \bibfnamefont {Jr.}}\ and\ \bibinfo {author} {\bibfnamefont {R.}~\bibnamefont {Ruffini}},\ }\bibfield  {title} {\bibinfo {title} {{Maximum mass of a neutron star}},\ }\href {https://doi.org/10.1103/PhysRevLett.32.324} {\bibfield  {journal} {\bibinfo  {journal} {Phys. Rev. Lett.}\ }\textbf {\bibinfo {volume} {32}},\ \bibinfo {pages} {324} (\bibinfo {year} {1974})}\BibitemShut {NoStop}%
\bibitem [{\citenamefont {Kalogera}\ and\ \citenamefont {Baym}(1996)}]{Kalogera:1996ci}%
  \BibitemOpen
  \bibfield  {author} {\bibinfo {author} {\bibfnamefont {V.}~\bibnamefont {Kalogera}}\ and\ \bibinfo {author} {\bibfnamefont {G.}~\bibnamefont {Baym}},\ }\bibfield  {title} {\bibinfo {title} {{The maximum mass of a neutron star}},\ }\href {https://doi.org/10.1086/310296} {\bibfield  {journal} {\bibinfo  {journal} {Astrophys. J. Lett.}\ }\textbf {\bibinfo {volume} {470}},\ \bibinfo {pages} {L61} (\bibinfo {year} {1996})},\ \Eprint {https://arxiv.org/abs/astro-ph/9608059} {arXiv:astro-ph/9608059} \BibitemShut {NoStop}%
\bibitem [{\citenamefont {Abbott}\ \emph {et~al.}(2023{\natexlab{a}})\citenamefont {Abbott} \emph {et~al.}}]{KAGRA:2021vkt}%
  \BibitemOpen
  \bibfield  {author} {\bibinfo {author} {\bibfnamefont {R.}~\bibnamefont {Abbott}} \emph {et~al.} (\bibinfo {collaboration} {KAGRA, VIRGO, LIGO Scientific}),\ }\bibfield  {title} {\bibinfo {title} {{GWTC-3: Compact Binary Coalescences Observed by LIGO and Virgo during the Second Part of the Third Observing Run}},\ }\href {https://doi.org/10.1103/PhysRevX.13.041039} {\bibfield  {journal} {\bibinfo  {journal} {Phys. Rev. X}\ }\textbf {\bibinfo {volume} {13}},\ \bibinfo {pages} {041039} (\bibinfo {year} {2023}{\natexlab{a}})},\ \Eprint {https://arxiv.org/abs/2111.03606} {arXiv:2111.03606 [gr-qc]} \BibitemShut {NoStop}%
\bibitem [{\citenamefont {Bailyn}\ \emph {et~al.}(1998)\citenamefont {Bailyn}, \citenamefont {Jain}, \citenamefont {Coppi},\ and\ \citenamefont {Orosz}}]{Bailyn:1997xt}%
  \BibitemOpen
  \bibfield  {author} {\bibinfo {author} {\bibfnamefont {C.~D.}\ \bibnamefont {Bailyn}}, \bibinfo {author} {\bibfnamefont {R.~K.}\ \bibnamefont {Jain}}, \bibinfo {author} {\bibfnamefont {P.}~\bibnamefont {Coppi}},\ and\ \bibinfo {author} {\bibfnamefont {J.~A.}\ \bibnamefont {Orosz}},\ }\bibfield  {title} {\bibinfo {title} {{The Mass distribution of stellar black holes}},\ }\href {https://doi.org/10.1086/305614} {\bibfield  {journal} {\bibinfo  {journal} {Astrophys. J.}\ }\textbf {\bibinfo {volume} {499}},\ \bibinfo {pages} {367} (\bibinfo {year} {1998})},\ \Eprint {https://arxiv.org/abs/astro-ph/9708032} {arXiv:astro-ph/9708032} \BibitemShut {NoStop}%
\bibitem [{\citenamefont {Ozel}\ \emph {et~al.}(2010)\citenamefont {Ozel}, \citenamefont {Psaltis}, \citenamefont {Narayan},\ and\ \citenamefont {McClintock}}]{Ozel:2010su}%
  \BibitemOpen
  \bibfield  {author} {\bibinfo {author} {\bibfnamefont {F.}~\bibnamefont {Ozel}}, \bibinfo {author} {\bibfnamefont {D.}~\bibnamefont {Psaltis}}, \bibinfo {author} {\bibfnamefont {R.}~\bibnamefont {Narayan}},\ and\ \bibinfo {author} {\bibfnamefont {J.~E.}\ \bibnamefont {McClintock}},\ }\bibfield  {title} {\bibinfo {title} {{The Black Hole Mass Distribution in the Galaxy}},\ }\href {https://doi.org/10.1088/0004-637X/725/2/1918} {\bibfield  {journal} {\bibinfo  {journal} {Astrophys. J.}\ }\textbf {\bibinfo {volume} {725}},\ \bibinfo {pages} {1918} (\bibinfo {year} {2010})},\ \Eprint {https://arxiv.org/abs/1006.2834} {arXiv:1006.2834 [astro-ph.GA]} \BibitemShut {NoStop}%
\bibitem [{\citenamefont {Farr}\ \emph {et~al.}(2011)\citenamefont {Farr}, \citenamefont {Sravan}, \citenamefont {Cantrell}, \citenamefont {Kreidberg}, \citenamefont {Bailyn}, \citenamefont {Mandel},\ and\ \citenamefont {Kalogera}}]{Farr:2010tu}%
  \BibitemOpen
  \bibfield  {author} {\bibinfo {author} {\bibfnamefont {W.~M.}\ \bibnamefont {Farr}}, \bibinfo {author} {\bibfnamefont {N.}~\bibnamefont {Sravan}}, \bibinfo {author} {\bibfnamefont {A.}~\bibnamefont {Cantrell}}, \bibinfo {author} {\bibfnamefont {L.}~\bibnamefont {Kreidberg}}, \bibinfo {author} {\bibfnamefont {C.~D.}\ \bibnamefont {Bailyn}}, \bibinfo {author} {\bibfnamefont {I.}~\bibnamefont {Mandel}},\ and\ \bibinfo {author} {\bibfnamefont {V.}~\bibnamefont {Kalogera}},\ }\bibfield  {title} {\bibinfo {title} {{The Mass Distribution of Stellar-Mass Black Holes}},\ }\href {https://doi.org/10.1088/0004-637X/741/2/103} {\bibfield  {journal} {\bibinfo  {journal} {Astrophys. J.}\ }\textbf {\bibinfo {volume} {741}},\ \bibinfo {pages} {103} (\bibinfo {year} {2011})},\ \Eprint {https://arxiv.org/abs/1011.1459} {arXiv:1011.1459 [astro-ph.GA]} \BibitemShut {NoStop}%
\bibitem [{\citenamefont {Kreidberg}\ \emph {et~al.}(2012)\citenamefont {Kreidberg}, \citenamefont {Bailyn}, \citenamefont {Farr},\ and\ \citenamefont {Kalogera}}]{Kreidberg:2012ud}%
  \BibitemOpen
  \bibfield  {author} {\bibinfo {author} {\bibfnamefont {L.}~\bibnamefont {Kreidberg}}, \bibinfo {author} {\bibfnamefont {C.~D.}\ \bibnamefont {Bailyn}}, \bibinfo {author} {\bibfnamefont {W.~M.}\ \bibnamefont {Farr}},\ and\ \bibinfo {author} {\bibfnamefont {V.}~\bibnamefont {Kalogera}},\ }\bibfield  {title} {\bibinfo {title} {{Mass Measurements of Black Holes in X-Ray Transients: Is There a Mass Gap?}},\ }\href {https://doi.org/10.1088/0004-637X/757/1/36} {\bibfield  {journal} {\bibinfo  {journal} {Astrophys. J.}\ }\textbf {\bibinfo {volume} {757}},\ \bibinfo {pages} {36} (\bibinfo {year} {2012})},\ \Eprint {https://arxiv.org/abs/1205.1805} {arXiv:1205.1805 [astro-ph.HE]} \BibitemShut {NoStop}%
\bibitem [{\citenamefont {Abbott}\ \emph {et~al.}(2020{\natexlab{b}})\citenamefont {Abbott} \emph {et~al.}}]{LIGOScientific:2020zkf}%
  \BibitemOpen
  \bibfield  {author} {\bibinfo {author} {\bibfnamefont {R.}~\bibnamefont {Abbott}} \emph {et~al.} (\bibinfo {collaboration} {LIGO Scientific, Virgo}),\ }\bibfield  {title} {\bibinfo {title} {{GW190814: Gravitational Waves from the Coalescence of a 23 Solar Mass Black Hole with a 2.6 Solar Mass Compact Object}},\ }\href {https://doi.org/10.3847/2041-8213/ab960f} {\bibfield  {journal} {\bibinfo  {journal} {Astrophys. J. Lett.}\ }\textbf {\bibinfo {volume} {896}},\ \bibinfo {pages} {L44} (\bibinfo {year} {2020}{\natexlab{b}})},\ \Eprint {https://arxiv.org/abs/2006.12611} {arXiv:2006.12611 [astro-ph.HE]} \BibitemShut {NoStop}%
\bibitem [{\citenamefont {Abbott}\ \emph {et~al.}(2024)\citenamefont {Abbott} \emph {et~al.}}]{LIGOScientific:2021usb}%
  \BibitemOpen
  \bibfield  {author} {\bibinfo {author} {\bibfnamefont {R.}~\bibnamefont {Abbott}} \emph {et~al.} (\bibinfo {collaboration} {LIGO Scientific, VIRGO}),\ }\bibfield  {title} {\bibinfo {title} {{GWTC-2.1: Deep extended catalog of compact binary coalescences observed by LIGO and Virgo during the first half of the third observing run}},\ }\href {https://doi.org/10.1103/PhysRevD.109.022001} {\bibfield  {journal} {\bibinfo  {journal} {Phys. Rev. D}\ }\textbf {\bibinfo {volume} {109}},\ \bibinfo {pages} {022001} (\bibinfo {year} {2024})},\ \Eprint {https://arxiv.org/abs/2108.01045} {arXiv:2108.01045 [gr-qc]} \BibitemShut {NoStop}%
\bibitem [{\citenamefont {Abbott}\ \emph {et~al.}(2023{\natexlab{b}})\citenamefont {Abbott} \emph {et~al.}}]{KAGRA:2021duu}%
  \BibitemOpen
  \bibfield  {author} {\bibinfo {author} {\bibfnamefont {R.}~\bibnamefont {Abbott}} \emph {et~al.} (\bibinfo {collaboration} {KAGRA, VIRGO, LIGO Scientific}),\ }\bibfield  {title} {\bibinfo {title} {{Population of Merging Compact Binaries Inferred Using Gravitational Waves through GWTC-3}},\ }\href {https://doi.org/10.1103/PhysRevX.13.011048} {\bibfield  {journal} {\bibinfo  {journal} {Phys. Rev. X}\ }\textbf {\bibinfo {volume} {13}},\ \bibinfo {pages} {011048} (\bibinfo {year} {2023}{\natexlab{b}})},\ \Eprint {https://arxiv.org/abs/2111.03634} {arXiv:2111.03634 [astro-ph.HE]} \BibitemShut {NoStop}%
\bibitem [{\citenamefont {Abac}\ \emph {et~al.}(2024{\natexlab{a}})\citenamefont {Abac} \emph {et~al.}}]{LIGOScientific:2024elc}%
  \BibitemOpen
  \bibfield  {author} {\bibinfo {author} {\bibfnamefont {A.~G.}\ \bibnamefont {Abac}} \emph {et~al.} (\bibinfo {collaboration} {LIGO Scientific, Virgo,, KAGRA, VIRGO}),\ }\bibfield  {title} {\bibinfo {title} {{Observation of Gravitational Waves from the Coalescence of a 2.5\textendash{}4.5 M $_{\odot}$ Compact Object and a Neutron Star}},\ }\href {https://doi.org/10.3847/2041-8213/ad5beb} {\bibfield  {journal} {\bibinfo  {journal} {Astrophys. J. Lett.}\ }\textbf {\bibinfo {volume} {970}},\ \bibinfo {pages} {L34} (\bibinfo {year} {2024}{\natexlab{a}})},\ \Eprint {https://arxiv.org/abs/2404.04248} {arXiv:2404.04248 [astro-ph.HE]} \BibitemShut {NoStop}%
\bibitem [{\citenamefont {Scheel}\ \emph {et~al.}(2015)\citenamefont {Scheel}, \citenamefont {Giesler}, \citenamefont {Hemberger}, \citenamefont {Lovelace}, \citenamefont {Kuper}, \citenamefont {Boyle}, \citenamefont {Szil\'agyi},\ and\ \citenamefont {Kidder}}]{Scheel:2014ina}%
  \BibitemOpen
  \bibfield  {author} {\bibinfo {author} {\bibfnamefont {M.~A.}\ \bibnamefont {Scheel}}, \bibinfo {author} {\bibfnamefont {M.}~\bibnamefont {Giesler}}, \bibinfo {author} {\bibfnamefont {D.~A.}\ \bibnamefont {Hemberger}}, \bibinfo {author} {\bibfnamefont {G.}~\bibnamefont {Lovelace}}, \bibinfo {author} {\bibfnamefont {K.}~\bibnamefont {Kuper}}, \bibinfo {author} {\bibfnamefont {M.}~\bibnamefont {Boyle}}, \bibinfo {author} {\bibfnamefont {B.}~\bibnamefont {Szil\'agyi}},\ and\ \bibinfo {author} {\bibfnamefont {L.~E.}\ \bibnamefont {Kidder}},\ }\bibfield  {title} {\bibinfo {title} {{Improved methods for simulating nearly extremal binary black holes}},\ }\href {https://doi.org/10.1088/0264-9381/32/10/105009} {\bibfield  {journal} {\bibinfo  {journal} {Class. Quant. Grav.}\ }\textbf {\bibinfo {volume} {32}},\ \bibinfo {pages} {105009} (\bibinfo {year} {2015})},\ \Eprint {https://arxiv.org/abs/1412.1803} {arXiv:1412.1803 [gr-qc]} \BibitemShut {NoStop}%
\bibitem [{\citenamefont {Boyle}\ \emph {et~al.}(2019)\citenamefont {Boyle} \emph {et~al.}}]{Boyle:2019kee}%
  \BibitemOpen
  \bibfield  {author} {\bibinfo {author} {\bibfnamefont {M.}~\bibnamefont {Boyle}} \emph {et~al.},\ }\bibfield  {title} {\bibinfo {title} {{The SXS Collaboration catalog of binary black hole simulations}},\ }\href {https://doi.org/10.1088/1361-6382/ab34e2} {\bibfield  {journal} {\bibinfo  {journal} {Class. Quant. Grav.}\ }\textbf {\bibinfo {volume} {36}},\ \bibinfo {pages} {195006} (\bibinfo {year} {2019})},\ \Eprint {https://arxiv.org/abs/1904.04831} {arXiv:1904.04831 [gr-qc]} \BibitemShut {NoStop}%
\bibitem [{\citenamefont {Scheel}\ \emph {et~al.}(2025)\citenamefont {Scheel} \emph {et~al.}}]{Scheel:2025jct}%
  \BibitemOpen
  \bibfield  {author} {\bibinfo {author} {\bibfnamefont {M.~A.}\ \bibnamefont {Scheel}} \emph {et~al.},\ }\bibfield  {title} {\bibinfo {title} {{The SXS Collaboration's third catalog of binary black hole simulations}},\ }\href@noop {} {\  (\bibinfo {year} {2025})},\ \Eprint {https://arxiv.org/abs/2505.13378} {arXiv:2505.13378 [gr-qc]} \BibitemShut {NoStop}%
\bibitem [{\citenamefont {Dietrich}\ \emph {et~al.}(2017)\citenamefont {Dietrich}, \citenamefont {Bernuzzi},\ and\ \citenamefont {Tichy}}]{Dietrich:2017aum}%
  \BibitemOpen
  \bibfield  {author} {\bibinfo {author} {\bibfnamefont {T.}~\bibnamefont {Dietrich}}, \bibinfo {author} {\bibfnamefont {S.}~\bibnamefont {Bernuzzi}},\ and\ \bibinfo {author} {\bibfnamefont {W.}~\bibnamefont {Tichy}},\ }\bibfield  {title} {\bibinfo {title} {{Closed-form tidal approximants for binary neutron star gravitational waveforms constructed from high-resolution numerical relativity simulations}},\ }\href {https://doi.org/10.1103/PhysRevD.96.121501} {\bibfield  {journal} {\bibinfo  {journal} {Phys. Rev. D}\ }\textbf {\bibinfo {volume} {96}},\ \bibinfo {pages} {121501} (\bibinfo {year} {2017})},\ \Eprint {https://arxiv.org/abs/1706.02969} {arXiv:1706.02969 [gr-qc]} \BibitemShut {NoStop}%
\bibitem [{\citenamefont {Dietrich}\ \emph {et~al.}(2019)\citenamefont {Dietrich}, \citenamefont {Samajdar}, \citenamefont {Khan}, \citenamefont {Johnson-McDaniel}, \citenamefont {Dudi},\ and\ \citenamefont {Tichy}}]{Dietrich:2019kaq}%
  \BibitemOpen
  \bibfield  {author} {\bibinfo {author} {\bibfnamefont {T.}~\bibnamefont {Dietrich}}, \bibinfo {author} {\bibfnamefont {A.}~\bibnamefont {Samajdar}}, \bibinfo {author} {\bibfnamefont {S.}~\bibnamefont {Khan}}, \bibinfo {author} {\bibfnamefont {N.~K.}\ \bibnamefont {Johnson-McDaniel}}, \bibinfo {author} {\bibfnamefont {R.}~\bibnamefont {Dudi}},\ and\ \bibinfo {author} {\bibfnamefont {W.}~\bibnamefont {Tichy}},\ }\bibfield  {title} {\bibinfo {title} {{Improving the NRTidal model for binary neutron star systems}},\ }\href {https://doi.org/10.1103/PhysRevD.100.044003} {\bibfield  {journal} {\bibinfo  {journal} {Phys. Rev. D}\ }\textbf {\bibinfo {volume} {100}},\ \bibinfo {pages} {044003} (\bibinfo {year} {2019})},\ \Eprint {https://arxiv.org/abs/1905.06011} {arXiv:1905.06011 [gr-qc]} \BibitemShut {NoStop}%
\bibitem [{\citenamefont {Abac}\ \emph {et~al.}(2024{\natexlab{b}})\citenamefont {Abac}, \citenamefont {Dietrich}, \citenamefont {Buonanno}, \citenamefont {Steinhoff},\ and\ \citenamefont {Ujevic}}]{Abac:2023ujg}%
  \BibitemOpen
  \bibfield  {author} {\bibinfo {author} {\bibfnamefont {A.}~\bibnamefont {Abac}}, \bibinfo {author} {\bibfnamefont {T.}~\bibnamefont {Dietrich}}, \bibinfo {author} {\bibfnamefont {A.}~\bibnamefont {Buonanno}}, \bibinfo {author} {\bibfnamefont {J.}~\bibnamefont {Steinhoff}},\ and\ \bibinfo {author} {\bibfnamefont {M.}~\bibnamefont {Ujevic}},\ }\bibfield  {title} {\bibinfo {title} {{New and robust gravitational-waveform model for high-mass-ratio binary neutron star systems with dynamical tidal effects}},\ }\href {https://doi.org/10.1103/PhysRevD.109.024062} {\bibfield  {journal} {\bibinfo  {journal} {Phys. Rev. D}\ }\textbf {\bibinfo {volume} {109}},\ \bibinfo {pages} {024062} (\bibinfo {year} {2024}{\natexlab{b}})},\ \Eprint {https://arxiv.org/abs/2311.07456} {arXiv:2311.07456 [gr-qc]} \BibitemShut {NoStop}%
\bibitem [{\citenamefont {P{\"u}rrer}\ and\ \citenamefont {Haster}(2020)}]{Purrer:2019jcp}%
  \BibitemOpen
  \bibfield  {author} {\bibinfo {author} {\bibfnamefont {M.}~\bibnamefont {P{\"u}rrer}}\ and\ \bibinfo {author} {\bibfnamefont {C.-J.}\ \bibnamefont {Haster}},\ }\bibfield  {title} {\bibinfo {title} {{Gravitational waveform accuracy requirements for future ground-based detectors}},\ }\href {https://doi.org/10.1103/PhysRevResearch.2.023151} {\bibfield  {journal} {\bibinfo  {journal} {Phys. Rev. Res.}\ }\textbf {\bibinfo {volume} {2}},\ \bibinfo {pages} {023151} (\bibinfo {year} {2020})},\ \Eprint {https://arxiv.org/abs/1912.10055} {arXiv:1912.10055 [gr-qc]} \BibitemShut {NoStop}%
\bibitem [{\citenamefont {Prasad}\ \emph {et~al.}(2020)\citenamefont {Prasad}, \citenamefont {Gupta}, \citenamefont {Bose}, \citenamefont {Krishnan},\ and\ \citenamefont {Schnetter}}]{Prasad:2020xgr}%
  \BibitemOpen
  \bibfield  {author} {\bibinfo {author} {\bibfnamefont {V.}~\bibnamefont {Prasad}}, \bibinfo {author} {\bibfnamefont {A.}~\bibnamefont {Gupta}}, \bibinfo {author} {\bibfnamefont {S.}~\bibnamefont {Bose}}, \bibinfo {author} {\bibfnamefont {B.}~\bibnamefont {Krishnan}},\ and\ \bibinfo {author} {\bibfnamefont {E.}~\bibnamefont {Schnetter}},\ }\bibfield  {title} {\bibinfo {title} {{News from horizons in binary black hole mergers}},\ }\href {https://doi.org/10.1103/PhysRevLett.125.121101} {\bibfield  {journal} {\bibinfo  {journal} {Phys. Rev. Lett.}\ }\textbf {\bibinfo {volume} {125}},\ \bibinfo {pages} {121101} (\bibinfo {year} {2020})},\ \Eprint {https://arxiv.org/abs/2003.06215} {arXiv:2003.06215 [gr-qc]} \BibitemShut {NoStop}%
\bibitem [{\citenamefont {Pook-Kolb}\ \emph {et~al.}(2020{\natexlab{a}})\citenamefont {Pook-Kolb}, \citenamefont {Birnholtz}, \citenamefont {Jaramillo}, \citenamefont {Krishnan},\ and\ \citenamefont {Schnetter}}]{Pook-Kolb:2020zhm}%
  \BibitemOpen
  \bibfield  {author} {\bibinfo {author} {\bibfnamefont {D.}~\bibnamefont {Pook-Kolb}}, \bibinfo {author} {\bibfnamefont {O.}~\bibnamefont {Birnholtz}}, \bibinfo {author} {\bibfnamefont {J.~L.}\ \bibnamefont {Jaramillo}}, \bibinfo {author} {\bibfnamefont {B.}~\bibnamefont {Krishnan}},\ and\ \bibinfo {author} {\bibfnamefont {E.}~\bibnamefont {Schnetter}},\ }\href@noop {} {\bibinfo {title} {{Horizons in a binary black hole merger I: Geometry and area increase}}} (\bibinfo {year} {2020}{\natexlab{a}}),\ \Eprint {https://arxiv.org/abs/2006.03939} {arXiv:2006.03939 [gr-qc]} \BibitemShut {NoStop}%
\bibitem [{\citenamefont {Pook-Kolb}\ \emph {et~al.}(2020{\natexlab{b}})\citenamefont {Pook-Kolb}, \citenamefont {Birnholtz}, \citenamefont {Jaramillo}, \citenamefont {Krishnan},\ and\ \citenamefont {Schnetter}}]{Pook-Kolb:2020jlr}%
  \BibitemOpen
  \bibfield  {author} {\bibinfo {author} {\bibfnamefont {D.}~\bibnamefont {Pook-Kolb}}, \bibinfo {author} {\bibfnamefont {O.}~\bibnamefont {Birnholtz}}, \bibinfo {author} {\bibfnamefont {J.~L.}\ \bibnamefont {Jaramillo}}, \bibinfo {author} {\bibfnamefont {B.}~\bibnamefont {Krishnan}},\ and\ \bibinfo {author} {\bibfnamefont {E.}~\bibnamefont {Schnetter}},\ }\href@noop {} {\bibinfo {title} {{Horizons in a binary black hole merger II: Fluxes, multipole moments and stability}}} (\bibinfo {year} {2020}{\natexlab{b}}),\ \Eprint {https://arxiv.org/abs/2006.03940} {arXiv:2006.03940 [gr-qc]} \BibitemShut {NoStop}%
\bibitem [{\citenamefont {Saketh}\ \emph {et~al.}(2023)\citenamefont {Saketh}, \citenamefont {Steinhoff}, \citenamefont {Vines},\ and\ \citenamefont {Buonanno}}]{Saketh:2022xjb}%
  \BibitemOpen
  \bibfield  {author} {\bibinfo {author} {\bibfnamefont {M.~V.~S.}\ \bibnamefont {Saketh}}, \bibinfo {author} {\bibfnamefont {J.}~\bibnamefont {Steinhoff}}, \bibinfo {author} {\bibfnamefont {J.}~\bibnamefont {Vines}},\ and\ \bibinfo {author} {\bibfnamefont {A.}~\bibnamefont {Buonanno}},\ }\bibfield  {title} {\bibinfo {title} {{Modeling horizon absorption in spinning binary black holes using effective worldline theory}},\ }\href {https://doi.org/10.1103/PhysRevD.107.084006} {\bibfield  {journal} {\bibinfo  {journal} {Phys. Rev. D}\ }\textbf {\bibinfo {volume} {107}},\ \bibinfo {pages} {084006} (\bibinfo {year} {2023})},\ \Eprint {https://arxiv.org/abs/2212.13095} {arXiv:2212.13095 [gr-qc]} \BibitemShut {NoStop}%
\bibitem [{\citenamefont {Blanchet}\ \emph {et~al.}(2023)\citenamefont {Blanchet}, \citenamefont {Faye}, \citenamefont {Henry}, \citenamefont {Larrouturou},\ and\ \citenamefont {Trestini}}]{Blanchet:2023bwj}%
  \BibitemOpen
  \bibfield  {author} {\bibinfo {author} {\bibfnamefont {L.}~\bibnamefont {Blanchet}}, \bibinfo {author} {\bibfnamefont {G.}~\bibnamefont {Faye}}, \bibinfo {author} {\bibfnamefont {Q.}~\bibnamefont {Henry}}, \bibinfo {author} {\bibfnamefont {F.}~\bibnamefont {Larrouturou}},\ and\ \bibinfo {author} {\bibfnamefont {D.}~\bibnamefont {Trestini}},\ }\bibfield  {title} {\bibinfo {title} {{Gravitational-Wave Phasing of Quasicircular Compact Binary Systems to the Fourth-and-a-Half Post-Newtonian Order}},\ }\href {https://doi.org/10.1103/PhysRevLett.131.121402} {\bibfield  {journal} {\bibinfo  {journal} {Phys. Rev. Lett.}\ }\textbf {\bibinfo {volume} {131}},\ \bibinfo {pages} {121402} (\bibinfo {year} {2023})},\ \Eprint {https://arxiv.org/abs/2304.11185} {arXiv:2304.11185 [gr-qc]} \BibitemShut {NoStop}%
\bibitem [{\citenamefont {Boyle}\ \emph {et~al.}(2008)\citenamefont {Boyle}, \citenamefont {Buonanno}, \citenamefont {Kidder}, \citenamefont {Mroue}, \citenamefont {Pan}, \citenamefont {Pfeiffer},\ and\ \citenamefont {Scheel}}]{Boyle:2008ge}%
  \BibitemOpen
  \bibfield  {author} {\bibinfo {author} {\bibfnamefont {M.}~\bibnamefont {Boyle}}, \bibinfo {author} {\bibfnamefont {A.}~\bibnamefont {Buonanno}}, \bibinfo {author} {\bibfnamefont {L.~E.}\ \bibnamefont {Kidder}}, \bibinfo {author} {\bibfnamefont {A.~H.}\ \bibnamefont {Mroue}}, \bibinfo {author} {\bibfnamefont {Y.}~\bibnamefont {Pan}}, \bibinfo {author} {\bibfnamefont {H.~P.}\ \bibnamefont {Pfeiffer}},\ and\ \bibinfo {author} {\bibfnamefont {M.~A.}\ \bibnamefont {Scheel}},\ }\bibfield  {title} {\bibinfo {title} {{High-accuracy numerical simulation of black-hole binaries: Computation of the gravitational-wave energy flux and comparisons with post-Newtonian approximants}},\ }\href {https://doi.org/10.1103/PhysRevD.78.104020} {\bibfield  {journal} {\bibinfo  {journal} {Phys. Rev. D}\ }\textbf {\bibinfo {volume} {78}},\ \bibinfo {pages} {104020} (\bibinfo {year} {2008})},\ \Eprint {https://arxiv.org/abs/0804.4184} {arXiv:0804.4184 [gr-qc]} \BibitemShut {NoStop}%
\bibitem [{\citenamefont {Gupta}\ \emph {et~al.}(2018)\citenamefont {Gupta}, \citenamefont {Krishnan}, \citenamefont {Nielsen},\ and\ \citenamefont {Schnetter}}]{Gupta:2018znn}%
  \BibitemOpen
  \bibfield  {author} {\bibinfo {author} {\bibfnamefont {A.}~\bibnamefont {Gupta}}, \bibinfo {author} {\bibfnamefont {B.}~\bibnamefont {Krishnan}}, \bibinfo {author} {\bibfnamefont {A.}~\bibnamefont {Nielsen}},\ and\ \bibinfo {author} {\bibfnamefont {E.}~\bibnamefont {Schnetter}},\ }\bibfield  {title} {\bibinfo {title} {{Dynamics of marginally trapped surfaces in a binary black hole merger: Growth and approach to equilibrium}},\ }\href {https://doi.org/10.1103/PhysRevD.97.084028} {\bibfield  {journal} {\bibinfo  {journal} {Phys. Rev. D}\ }\textbf {\bibinfo {volume} {97}},\ \bibinfo {pages} {084028} (\bibinfo {year} {2018})},\ \Eprint {https://arxiv.org/abs/1801.07048} {arXiv:1801.07048 [gr-qc]} \BibitemShut {NoStop}%
\bibitem [{\citenamefont {Tichy}\ \emph {et~al.}(2000)\citenamefont {Tichy}, \citenamefont {Flanagan},\ and\ \citenamefont {Poisson}}]{Tichy:1999pv}%
  \BibitemOpen
  \bibfield  {author} {\bibinfo {author} {\bibfnamefont {W.}~\bibnamefont {Tichy}}, \bibinfo {author} {\bibfnamefont {E.~E.}\ \bibnamefont {Flanagan}},\ and\ \bibinfo {author} {\bibfnamefont {E.}~\bibnamefont {Poisson}},\ }\bibfield  {title} {\bibinfo {title} {{Can the postNewtonian gravitational wave form of an inspiraling binary be improved by solving the energy balance equation numerically?}},\ }\href {https://doi.org/10.1103/PhysRevD.61.104015} {\bibfield  {journal} {\bibinfo  {journal} {Phys. Rev.}\ }\textbf {\bibinfo {volume} {D61}},\ \bibinfo {pages} {104015} (\bibinfo {year} {2000})},\ \Eprint {https://arxiv.org/abs/gr-qc/9912075} {arXiv:gr-qc/9912075 [gr-qc]} \BibitemShut {NoStop}%
\bibitem [{\citenamefont {Healy}\ \emph {et~al.}(2017)\citenamefont {Healy}, \citenamefont {Lousto},\ and\ \citenamefont {Zlochower}}]{Healy:2017mvh}%
  \BibitemOpen
  \bibfield  {author} {\bibinfo {author} {\bibfnamefont {J.}~\bibnamefont {Healy}}, \bibinfo {author} {\bibfnamefont {C.~O.}\ \bibnamefont {Lousto}},\ and\ \bibinfo {author} {\bibfnamefont {Y.}~\bibnamefont {Zlochower}},\ }\bibfield  {title} {\bibinfo {title} {{Nonspinning binary black hole merger scenario revisited}},\ }\href {https://doi.org/10.1103/PhysRevD.96.024031} {\bibfield  {journal} {\bibinfo  {journal} {Phys. Rev. D}\ }\textbf {\bibinfo {volume} {96}},\ \bibinfo {pages} {024031} (\bibinfo {year} {2017})},\ \Eprint {https://arxiv.org/abs/1705.07034} {arXiv:1705.07034 [gr-qc]} \BibitemShut {NoStop}%
\bibitem [{\citenamefont {Mukherjee}\ \emph {et~al.}(2024)\citenamefont {Mukherjee}, \citenamefont {Phukon}, \citenamefont {Datta},\ and\ \citenamefont {Bose}}]{Mukherjee:2023pge}%
  \BibitemOpen
  \bibfield  {author} {\bibinfo {author} {\bibfnamefont {S.}~\bibnamefont {Mukherjee}}, \bibinfo {author} {\bibfnamefont {K.~S.}\ \bibnamefont {Phukon}}, \bibinfo {author} {\bibfnamefont {S.}~\bibnamefont {Datta}},\ and\ \bibinfo {author} {\bibfnamefont {S.}~\bibnamefont {Bose}},\ }\bibfield  {title} {\bibinfo {title} {{Phenomenological gravitational waveform model of binary black holes incorporating horizon fluxes}},\ }\href {https://doi.org/10.1103/PhysRevD.110.124027} {\bibfield  {journal} {\bibinfo  {journal} {Phys. Rev. D}\ }\textbf {\bibinfo {volume} {110}},\ \bibinfo {pages} {124027} (\bibinfo {year} {2024})},\ \Eprint {https://arxiv.org/abs/2311.17554} {arXiv:2311.17554 [gr-qc]} \BibitemShut {NoStop}%
\bibitem [{\citenamefont {Ghosh}\ and\ \citenamefont {Hannam}(2025)}]{Ghosh:2025wex}%
  \BibitemOpen
  \bibfield  {author} {\bibinfo {author} {\bibfnamefont {S.}~\bibnamefont {Ghosh}}\ and\ \bibinfo {author} {\bibfnamefont {M.}~\bibnamefont {Hannam}},\ }\bibfield  {title} {\bibinfo {title} {{On the Identification of Exotic Compact Binaries with Gravitational Waves: a Phenomenological approach}},\ }\href@noop {} {\  (\bibinfo {year} {2025})},\ \Eprint {https://arxiv.org/abs/2505.16380} {arXiv:2505.16380 [gr-qc]} \BibitemShut {NoStop}%
\bibitem [{\citenamefont {Cardoso}\ and\ \citenamefont {Pani}(2017)}]{cardoso:2017cqb}%
  \BibitemOpen
  \bibfield  {author} {\bibinfo {author} {\bibfnamefont {V.}~\bibnamefont {Cardoso}}\ and\ \bibinfo {author} {\bibfnamefont {P.}~\bibnamefont {Pani}},\ }\bibfield  {title} {\bibinfo {title} {{Tests for the existence of black holes through gravitational wave echoes}},\ }\href {https://doi.org/10.1038/s41550-017-0225-y} {\bibfield  {journal} {\bibinfo  {journal} {Nature Astron.}\ }\textbf {\bibinfo {volume} {1}},\ \bibinfo {pages} {586} (\bibinfo {year} {2017})},\ \Eprint {https://arxiv.org/abs/1709.01525} {arXiv:1709.01525 [gr-qc]} \BibitemShut {NoStop}%
\bibitem [{\citenamefont {Conklin}\ and\ \citenamefont {Holdom}(2019)}]{Conklin:2019fcs}%
  \BibitemOpen
  \bibfield  {author} {\bibinfo {author} {\bibfnamefont {R.~S.}\ \bibnamefont {Conklin}}\ and\ \bibinfo {author} {\bibfnamefont {B.}~\bibnamefont {Holdom}},\ }\bibfield  {title} {\bibinfo {title} {{Gravitational wave echo spectra}},\ }\href {https://doi.org/10.1103/PhysRevD.100.124030} {\bibfield  {journal} {\bibinfo  {journal} {Phys. Rev. D}\ }\textbf {\bibinfo {volume} {100}},\ \bibinfo {pages} {124030} (\bibinfo {year} {2019})},\ \Eprint {https://arxiv.org/abs/1905.09370} {arXiv:1905.09370 [gr-qc]} \BibitemShut {NoStop}%
\bibitem [{\citenamefont {Dietrich}\ \emph {et~al.}(2021)\citenamefont {Dietrich}, \citenamefont {Hinderer},\ and\ \citenamefont {Samajdar}}]{Dietrich:2020eud}%
  \BibitemOpen
  \bibfield  {author} {\bibinfo {author} {\bibfnamefont {T.}~\bibnamefont {Dietrich}}, \bibinfo {author} {\bibfnamefont {T.}~\bibnamefont {Hinderer}},\ and\ \bibinfo {author} {\bibfnamefont {A.}~\bibnamefont {Samajdar}},\ }\bibfield  {title} {\bibinfo {title} {{Interpreting Binary Neutron Star Mergers: Describing the Binary Neutron Star Dynamics, Modelling Gravitational Waveforms, and Analyzing Detections}},\ }\href {https://doi.org/10.1007/s10714-020-02751-6} {\bibfield  {journal} {\bibinfo  {journal} {Gen. Rel. Grav.}\ }\textbf {\bibinfo {volume} {53}},\ \bibinfo {pages} {27} (\bibinfo {year} {2021})},\ \Eprint {https://arxiv.org/abs/2004.02527} {arXiv:2004.02527 [gr-qc]} \BibitemShut {NoStop}%
\bibitem [{\citenamefont {Binnington}\ and\ \citenamefont {Poisson}(2009)}]{Binnington:2009bb}%
  \BibitemOpen
  \bibfield  {author} {\bibinfo {author} {\bibfnamefont {T.}~\bibnamefont {Binnington}}\ and\ \bibinfo {author} {\bibfnamefont {E.}~\bibnamefont {Poisson}},\ }\bibfield  {title} {\bibinfo {title} {{Relativistic theory of tidal Love numbers}},\ }\href {https://doi.org/10.1103/PhysRevD.80.084018} {\bibfield  {journal} {\bibinfo  {journal} {Phys. Rev.}\ }\textbf {\bibinfo {volume} {D80}},\ \bibinfo {pages} {084018} (\bibinfo {year} {2009})},\ \Eprint {https://arxiv.org/abs/0906.1366} {arXiv:0906.1366 [gr-qc]} \BibitemShut {NoStop}%
\bibitem [{\citenamefont {Ghosh}\ \emph {et~al.}(2024)\citenamefont {Ghosh}, \citenamefont {Pradhan},\ and\ \citenamefont {Chatterjee}}]{Ghosh:2023vrx}%
  \BibitemOpen
  \bibfield  {author} {\bibinfo {author} {\bibfnamefont {S.}~\bibnamefont {Ghosh}}, \bibinfo {author} {\bibfnamefont {B.~K.}\ \bibnamefont {Pradhan}},\ and\ \bibinfo {author} {\bibfnamefont {D.}~\bibnamefont {Chatterjee}},\ }\bibfield  {title} {\bibinfo {title} {{Tidal heating as a direct probe of strangeness inside neutron stars}},\ }\href {https://doi.org/10.1103/PhysRevD.109.103036} {\bibfield  {journal} {\bibinfo  {journal} {Phys. Rev. D}\ }\textbf {\bibinfo {volume} {109}},\ \bibinfo {pages} {103036} (\bibinfo {year} {2024})},\ \Eprint {https://arxiv.org/abs/2306.14737} {arXiv:2306.14737 [gr-qc]} \BibitemShut {NoStop}%
\bibitem [{\citenamefont {Ghosh}\ \emph {et~al.}(2025)\citenamefont {Ghosh}, \citenamefont {Mukherjee}, \citenamefont {Bose},\ and\ \citenamefont {Chatterjee}}]{Ghosh:2025glz}%
  \BibitemOpen
  \bibfield  {author} {\bibinfo {author} {\bibfnamefont {S.}~\bibnamefont {Ghosh}}, \bibinfo {author} {\bibfnamefont {S.}~\bibnamefont {Mukherjee}}, \bibinfo {author} {\bibfnamefont {S.}~\bibnamefont {Bose}},\ and\ \bibinfo {author} {\bibfnamefont {D.}~\bibnamefont {Chatterjee}},\ }\bibfield  {title} {\bibinfo {title} {{Tidal dissipation in binary neutron star inspiral : Bias study and modeling of frequency domain phase}},\ }\href@noop {} {\  (\bibinfo {year} {2025})},\ \Eprint {https://arxiv.org/abs/2503.14606} {arXiv:2503.14606 [gr-qc]} \BibitemShut {NoStop}%
\bibitem [{\citenamefont {Saketh}\ \emph {et~al.}(2024)\citenamefont {Saketh}, \citenamefont {Zhou}, \citenamefont {Ghosh}, \citenamefont {Steinhoff},\ and\ \citenamefont {Chatterjee}}]{Saketh:2024juq}%
  \BibitemOpen
  \bibfield  {author} {\bibinfo {author} {\bibfnamefont {M.~V.~S.}\ \bibnamefont {Saketh}}, \bibinfo {author} {\bibfnamefont {Z.}~\bibnamefont {Zhou}}, \bibinfo {author} {\bibfnamefont {S.}~\bibnamefont {Ghosh}}, \bibinfo {author} {\bibfnamefont {J.}~\bibnamefont {Steinhoff}},\ and\ \bibinfo {author} {\bibfnamefont {D.}~\bibnamefont {Chatterjee}},\ }\bibfield  {title} {\bibinfo {title} {{Investigating tidal heating in neutron stars via gravitational Raman scattering}},\ }\href {https://doi.org/10.1103/PhysRevD.110.103001} {\bibfield  {journal} {\bibinfo  {journal} {Phys. Rev. D}\ }\textbf {\bibinfo {volume} {110}},\ \bibinfo {pages} {103001} (\bibinfo {year} {2024})},\ \Eprint {https://arxiv.org/abs/2407.08327} {arXiv:2407.08327 [gr-qc]} \BibitemShut {NoStop}%
\bibitem [{\citenamefont {Hegade K.~R.}\ \emph {et~al.}(2024)\citenamefont {Hegade K.~R.}, \citenamefont {Ripley},\ and\ \citenamefont {Yunes}}]{HegadeKR:2024slr}%
  \BibitemOpen
  \bibfield  {author} {\bibinfo {author} {\bibfnamefont {A.}~\bibnamefont {Hegade K.~R.}}, \bibinfo {author} {\bibfnamefont {J.~L.}\ \bibnamefont {Ripley}},\ and\ \bibinfo {author} {\bibfnamefont {N.}~\bibnamefont {Yunes}},\ }\bibfield  {title} {\bibinfo {title} {{Dissipative tidal effects to next-to-leading order and constraints on the dissipative tidal deformability using gravitational wave data}},\ }\href {https://doi.org/10.1103/PhysRevD.110.044041} {\bibfield  {journal} {\bibinfo  {journal} {Phys. Rev. D}\ }\textbf {\bibinfo {volume} {110}},\ \bibinfo {pages} {044041} (\bibinfo {year} {2024})},\ \Eprint {https://arxiv.org/abs/2407.02584} {arXiv:2407.02584 [gr-qc]} \BibitemShut {NoStop}%
\bibitem [{\citenamefont {Pratten}\ \emph {et~al.}(2020)\citenamefont {Pratten}, \citenamefont {Schmidt},\ and\ \citenamefont {Hinderer}}]{Pratten:2019sed}%
  \BibitemOpen
  \bibfield  {author} {\bibinfo {author} {\bibfnamefont {G.}~\bibnamefont {Pratten}}, \bibinfo {author} {\bibfnamefont {P.}~\bibnamefont {Schmidt}},\ and\ \bibinfo {author} {\bibfnamefont {T.}~\bibnamefont {Hinderer}},\ }\bibfield  {title} {\bibinfo {title} {{Gravitational-Wave Asteroseismology with Fundamental Modes from Compact Binary Inspirals}},\ }\href {https://doi.org/10.1038/s41467-020-15984-5} {\bibfield  {journal} {\bibinfo  {journal} {Nature Commun.}\ }\textbf {\bibinfo {volume} {11}},\ \bibinfo {pages} {2553} (\bibinfo {year} {2020})},\ \Eprint {https://arxiv.org/abs/1905.00817} {arXiv:1905.00817 [gr-qc]} \BibitemShut {NoStop}%
\bibitem [{\citenamefont {Biswas}\ and\ \citenamefont {Datta}(2022)}]{Biswas:2021paf}%
  \BibitemOpen
  \bibfield  {author} {\bibinfo {author} {\bibfnamefont {B.}~\bibnamefont {Biswas}}\ and\ \bibinfo {author} {\bibfnamefont {S.}~\bibnamefont {Datta}},\ }\bibfield  {title} {\bibinfo {title} {{Constraining neutron star properties with a new equation of state insensitive approach}},\ }\href {https://doi.org/10.1103/PhysRevD.106.043012} {\bibfield  {journal} {\bibinfo  {journal} {Phys. Rev. D}\ }\textbf {\bibinfo {volume} {106}},\ \bibinfo {pages} {043012} (\bibinfo {year} {2022})},\ \Eprint {https://arxiv.org/abs/2112.10824} {arXiv:2112.10824 [astro-ph.HE]} \BibitemShut {NoStop}%
\bibitem [{\citenamefont {{LIGO Scientific Collaboration}}\ \emph {et~al.}(2018)\citenamefont {{LIGO Scientific Collaboration}}, \citenamefont {{Virgo Collaboration}},\ and\ \citenamefont {{KAGRA Collaboration}}}]{lalsuite}%
  \BibitemOpen
  \bibfield  {author} {\bibinfo {author} {\bibnamefont {{LIGO Scientific Collaboration}}}, \bibinfo {author} {\bibnamefont {{Virgo Collaboration}}},\ and\ \bibinfo {author} {\bibnamefont {{KAGRA Collaboration}}},\ }\href {https://doi.org/10.7935/GT1W-FZ16} {\bibinfo {title} {{LVK} {A}lgorithm {L}ibrary - {LALS}uite}},\ \bibinfo {howpublished} {Free software (GPL)} (\bibinfo {year} {2018})\BibitemShut {NoStop}%
\bibitem [{\citenamefont {Evans}\ \emph {et~al.}(2021)\citenamefont {Evans} \emph {et~al.}}]{Evans:2021gyd}%
  \BibitemOpen
  \bibfield  {author} {\bibinfo {author} {\bibfnamefont {M.}~\bibnamefont {Evans}} \emph {et~al.},\ }\bibfield  {title} {\bibinfo {title} {{A Horizon Study for Cosmic Explorer: Science, Observatories, and Community}},\ }\href@noop {} {\  (\bibinfo {year} {2021})},\ \Eprint {https://arxiv.org/abs/2109.09882} {arXiv:2109.09882 [astro-ph.IM]} \BibitemShut {NoStop}%
\bibitem [{\citenamefont {Harris}\ \emph {et~al.}(2020)\citenamefont {Harris} \emph {et~al.}}]{Harris:2020xlr}%
  \BibitemOpen
  \bibfield  {author} {\bibinfo {author} {\bibfnamefont {C.~R.}\ \bibnamefont {Harris}} \emph {et~al.},\ }\bibfield  {title} {\bibinfo {title} {{Array programming with NumPy}},\ }\href {https://doi.org/10.1038/s41586-020-2649-2} {\bibfield  {journal} {\bibinfo  {journal} {Nature}\ }\textbf {\bibinfo {volume} {585}},\ \bibinfo {pages} {357} (\bibinfo {year} {2020})},\ \Eprint {https://arxiv.org/abs/2006.10256} {arXiv:2006.10256 [cs.MS]} \BibitemShut {NoStop}%
\bibitem [{\citenamefont {Hunter}(2007)}]{Hunter:2007ouj}%
  \BibitemOpen
  \bibfield  {author} {\bibinfo {author} {\bibfnamefont {J.~D.}\ \bibnamefont {Hunter}},\ }\bibfield  {title} {\bibinfo {title} {{Matplotlib: A 2D Graphics Environment}},\ }\href {https://doi.org/10.1109/MCSE.2007.55} {\bibfield  {journal} {\bibinfo  {journal} {Comput. Sci. Eng.}\ }\textbf {\bibinfo {volume} {9}},\ \bibinfo {pages} {90} (\bibinfo {year} {2007})}\BibitemShut {NoStop}%
\bibitem [{\citenamefont {Virtanen}\ \emph {et~al.}(2020)\citenamefont {Virtanen} \emph {et~al.}}]{2020SciPy-NMeth}%
  \BibitemOpen
  \bibfield  {author} {\bibinfo {author} {\bibfnamefont {P.}~\bibnamefont {Virtanen}} \emph {et~al.},\ }\bibfield  {title} {\bibinfo {title} {{{SciPy} 1.0: Fundamental Algorithms for Scientific Computing in Python}},\ }\href {https://doi.org/10.1038/s41592-019-0686-2} {\bibfield  {journal} {\bibinfo  {journal} {Nat. Methods}\ }\textbf {\bibinfo {volume} {17}},\ \bibinfo {pages} {261} (\bibinfo {year} {2020})},\ \Eprint {https://arxiv.org/abs/1907.10121} {arXiv:1907.10121 [cs.MS]} \BibitemShut {NoStop}%
\bibitem [{\citenamefont {Boyle}\ and\ \citenamefont {Scheel}(2024)}]{boyle_2024_13714441}%
  \BibitemOpen
  \bibfield  {author} {\bibinfo {author} {\bibfnamefont {M.}~\bibnamefont {Boyle}}\ and\ \bibinfo {author} {\bibfnamefont {M.}~\bibnamefont {Scheel}},\ }\href {https://doi.org/10.5281/zenodo.13714441} {\bibinfo {title} {The sxs package}} (\bibinfo {year} {2024})\BibitemShut {NoStop}%
\end{thebibliography}
\end{document}